\documentclass{aa}  

\usepackage{graphicx}
\usepackage{txfonts}
\usepackage{lipsum}
\usepackage{subcaption}
\usepackage{lscape}
\usepackage{placeins} 
\usepackage{float}
\usepackage{verbatim}
\usepackage{xcolor}
\usepackage{amssymb}
\usepackage{xspace}
\usepackage{hyperref}
\usepackage[normalem]{ulem}
\hypersetup{
    colorlinks=true,
    linkcolor=teal,
    filecolor=magenta, 
    citecolor=teal,
    urlcolor=cyan
    }

\newcommand{\rh}[1]{{#1}}
\newcommand{\thethr}{\textsc{The Three Hundred}\xspace}
\newcommand{\theth}{\textsc{The300}\xspace}
\newcommand{\pymsz}{\textsc{PYMSZ}\xspace}

\newcommand{\hMpc}{{\ifmmode{\,h^{-1}{\rm Mpc}}\else{$h^{-1}$Mpc}\fi}}
\newcommand{\hkpc}{{\ifmmode{\,h^{-1}{\rm kpc}}\else{$h^{-1}$kpc}\fi}}
\newcommand{\hMsun}{{\ifmmode{\,h^{-1}{\rm {M_{\odot}}}}\else{$h^{-1}{\rm{M_{\odot}}}$}\fi}}
\newcommand{\Msun}{M$_\odot$}
\newcommand{\Mstar}{{\ifmmode{\,M_{*}}\else{$M_{*}$}\fi}}
\newcommand{\Mhalo}{{\ifmmode{\,M_{\rm halo}}\else{$M_{\rm halo}$}\fi}}

\newcommand{\caesar}{\textsc{Caesar}\xspace}

\newcommand{\gadgetx}{\textsc{Gadget-X}\xspace}
\newcommand{\simba}{\textsc{Simba}\xspace}
\newcommand{\gizmo}{\textsc{Gizmo}\xspace}

\newcommand{\simbac}{\textsc{Simba-C}\xspace}

\newcommand{\dv}[1]{\mathrm{d} #1}

\begin{document}

\title{Here, There and Everywhere: How AGN jets affect galaxy cluster environments}

\author{Isaac Rosenberg\inst{1, 2, 3, 4}\fnmsep\thanks{e-mail: isaacjerome.rosenberg@students.uniroma2.eu}
\and
{Martine Lokken}\inst{5}
\and
{Renée Hložek}\inst{3, 4}
\and
{Weiguang Cui}\inst{7,8,10}
\and
{Sara Santoni}\inst{6,7}
\and
{Romeel Dav\'e}\inst{8,9}}
\institute{
Department of Physics, University of Rome Tor Vergata, 
via della Ricerca Scientifica 1, 00133 Rome, Italy
\and
Department of Astronomy, University of Belgrade - Faculty of Mathematics, Studentski trg 16, 11000 Belgrade, Serbia
\and
David A. Dunlap Department of Astronomy and Astrophysics, 50 St George St, University of Toronto, Toronto, ON M5S 3H8, Canada
\and
Department of Physics, 60 St George St, University of Toronto, Toronto, ON M5S 3H8, Canada
\and
Institut de Fisica d’Altes Energies (IFAE), The Barcelona Institute of Science and Technology, Campus UAB, 08193 Bellaterra (Barcelona) Spain
\and 
Dipartimento di Fisica, Sapienza Università di Roma, Piazzale Aldo Moro 5, I-00185 Rome, Italy
\and 
Departamento de Física Teórica, Facultad de Ciencias, Universidad Autónoma de Madrid, Modulo 8, E-28049 Madrid, Spain
\and
Institute for Astronomy, University of Edinburgh, EH8 9SQ, Edinburgh, U.K.
\and
Department of Physics and Astronomy, University of the Western Cape, Bellville 7535, Cape Town, South Africa
\and
Centro de Investigación Avanzada en Física Fundamental (CIAFF), Facultad de Ciencias, Universidad Autónoma de Madrid, 28049 Madrid, Spain}

\abstract
{Active galactic nuclei (AGN) feedback via black hole-driven jets and winds plays a key role in redistributing matter across megaparsec scales. However, the implementation of jet feedback in cosmological simulations remains highly prescriptive, leading to uncertainties in the predicted state of the Warm-Hot Intergalactic Medium (WHIM) and other extragalactic observables.}
{We investigate how variations in AGN jet velocity, orientation, and delayed hydrodynamic coupling impact the thermodynamic state of gas surrounding galaxy clusters. We aim to identify observational signatures in the thermal Sunyaev-Zel’dovich (tSZ) effect and galaxy properties that can constrain these models.}
{We employ zoom-in hydrodynamic simulations centered on three galaxy clusters from \thethr, utilizing the \simbac model and various feedback parameters. We use filament-finding algorithms and oriented stacking to probe the effect on large-scale structure and compare our results to observational data for brightest cluster galaxies from eRASS1, black hole-halo mass relations and baryonic mass fractions.}
{We find that jet velocity is the dominant parameter of those we tested in heating low-density environments at $z > 1$. At lower redshifts, higher velocity jets efficiently quench star formation, expel baryonic matter, and prevent black hole growth. While the tSZ signal within the central cluster and surrounding filaments is only mildly affected ($\sim10\%$), the signal in under-dense regions outside filaments is enhanced by $\sim100\%$. Modifying the decoupling time and jet orientation have subtler, non-uniform effects on the gas on larger scales.}
{In this case study, high velocity AGN jets provide the best match to galaxy properties from eRASS1 and other X-ray surveys. We find that hot gas in low-density regimes is a sensitive probe of AGN feedback. Future high-resolution tSZ surveys like the Simons Observatory and spectral-distortion experiments like FOSSIL have the potential to probe the thermal state of the gas outside clusters to distinguish between these feedback models.}


   \maketitle

\nolinenumbers


\section{Introduction} 

Active galactic nuclei (AGN) are supermassive black holes (BHs) whose accretion disks emit large amounts of energy across the electromagnetic spectrum. Observational evidence suggests that some AGN launch relativistic collimated jets of charged particles and radiation. The jets can travel all the way to a few megaparsecs (Mpc), sometimes terminating in observable radio lobes \citep{Heckman_2014, Clarke_2017,Dabhade_2023, Oei_2024, 2025MNRAS.538.1628B}.

By pumping gas from halo centers to such large distance scales, `AGN feedback' can shape the properties of galaxies and their surroundings. AGN have been demonstrated to drive quenching in both observations and simulations, removing cold gas that galaxies would otherwise use for star formation, impacting the stellar mass function, colors, and star formation histories of galaxies \citep{CanoDiaz2012A&A...537L...8C, Schaye_2014, Vogelsberger_2014, Dav__2019, Cui2021, Piotrowska_2021, Irodotou_2022,Bigwood_2025,rasia_2025, Iyer2025ApJ...994..174I}. Properties of the circumgalactic medium such as surface brightness and density are also impacted by varying levels of AGN feedback \citep{grayson2025hotcircumgalacticmediumstacked}.

The impacts on larger scales are more difficult to assess, as intergalactic medium (IGM) gas is diffuse and therefore challenging to observe. Simulations have shown that AGN feedback is needed to match observations in the $\textrm{Ly}\alpha$ forest \citep{Christiansen_2020, Tillman_2023, Tillman_2025}. Recent kinetic Sunyaev-Zel'dovich (kSZ) measurements suggest that gas is distributed further beyond halos than previously thought \citep{Bigwood2024MNRAS.534..655B, Hadzhiyska2025PhRvD.112h3509H}, and various simulations suggest that AGN, in particular jets, could be responsible \citep{Biffi_2018,sorini2022MNRAS.516..883S}. Meanwhile, the discovery of many giant radio jets in observational data \citep{oei2023A&A...672A.163O} has given more weight to the possibility of this connection. Future observations of diffuse gas with the kSZ and the thermal SZ (tSZ), X-ray, and other probes like absorption lines and fast radio burst dispersion measures could continue to develop this picture. Understanding the motion and redistribution of the baryons through large-scale structure over cosmic time is a key part of verifying or breaking the standard models of cosmology and structure formation \citep[e.g., solving the `missing baryon problem', see][for a recent review]{baryonprob_review}. Large-scale impacts are also a key unknown currently limiting precision cosmology with gravitational lensing and other probes \citep{Harnois-Deraps2015}.

\begin{table*}[htbp!]
\centering
\caption{Summary of the simulation runs across different regions, varying jet velocity caps, decoupling times, and jet directions. The name of the runs, as they will be referred to in the text and plots, is shown in the right-most column.}
\begin{tabular}{cccccc}
\hline
\hline
Region & Jet velocity cap  & Decoupling Scale ($\alpha$)  & Jet Direction & Notes & Name\\
 & [km\,s$^{-1}$] & $t_{\mathrm{Decoupling}} = \alpha~t_{\mathrm{Hubble}}$ &  &  & \\
\hline
A & $7,000$   & $10^{-4}$ & Angular Momentum & Default setup & Baseline \\
A & $35,000$   & $10^{-4}$ & Angular Momentum & Higher $v_{\rm max}$ & Fast jets\\

A & N/A    & $10^{-4}$ & Angular Momentum & No jets & No jets\\
A & $7,000$    & $10^{-4}$ & Randomized       & Random jet orientation & Random\\

B & N/A   & $10^{-4}$ & Angular Momentum & No jets & \\
B & $7,000$    & $10^{-4}$ & Angular Momentum & Default setup & \\
B & $35,000$   & $10^{-4}$ & Angular Momentum & Higher $v_{\rm max}$ & \\
C & N/A   & $10^{-4}$ & Angular Momentum & No jets & \\
C & $7,000$    & $10^{-4}$ & Angular Momentum & Default setup & \\
C & $35,000$   & $10^{-4}$ & Angular Momentum & Higher $v_{\rm max}$ & \\
C & $7,000$    & $2 \cdot 10^{-3}$ & Angular Momentum & Longer decoupling time & Late coupling \\
\hline
\end{tabular}
\label{tab:runs_summary}
\end{table*}

\begin{figure*}
    \centering
    \includegraphics[width=0.85\linewidth]{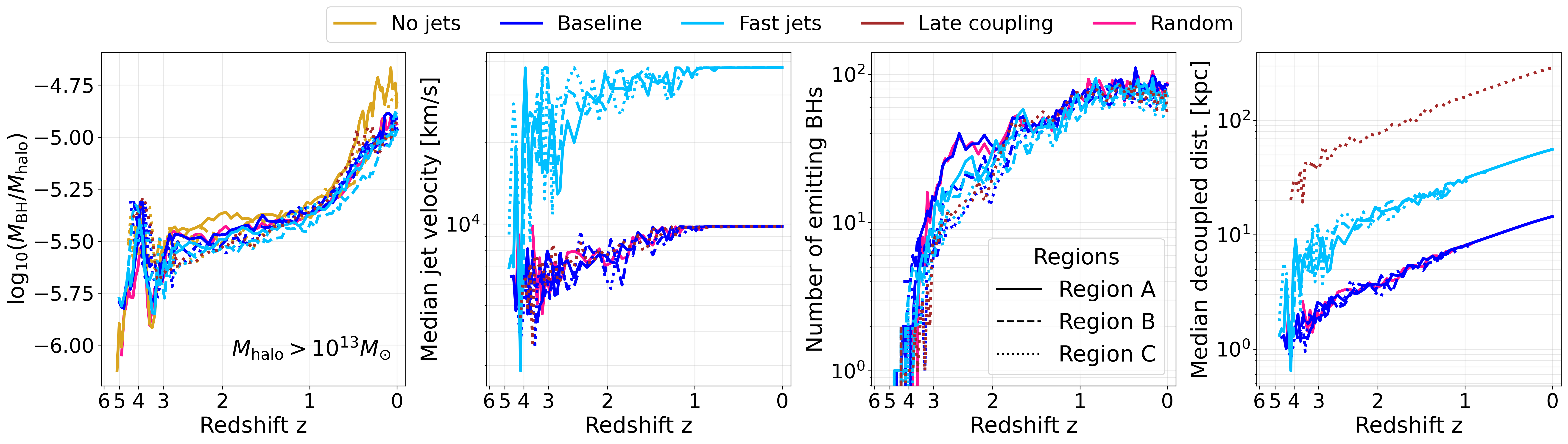}
    \caption{Various jet properties as a function of redshift for all of the runs. The first column is the black hole mass fraction for halos above $10^{13}\, \mathrm{M_{\odot}}$. The second column is the median active jet velocities. The third column is the number of active jets. The fourth column is the median distance at which jets emitted at that time would be recoupled. AGN feedback models are denoted by different line colors, while simulation regions are denoted by line style.
    \label{fig:jet_props}}
\end{figure*}

To inform future observational campaigns to constrain AGN impacts, we need simulations of the formation and evolution of cosmic structures which incorporate detailed gas processes. Hydrodynamic (hydro) simulations offer a controlled laboratory to study how different physical processes shape large-scale morphology.
However, due to computational limitations, it is impossible to simultaneously simulate the scales responsible for AGN energy generation (sub-kpc accretion disks surrounding BHs) as well as the multi-Mpc scales of structure formation. Therefore, cosmological hydro simulations use prescriptive models to represent the effects that emerge from sub-grid processes. These sub-grid models are difficult to constrain empirically and are typically tuned to reproduce particular observations such as correlations between quiescent galaxies and properties including BH mass, halo mass, position within halo, gas fraction, and cluster scaling relations from X-ray observations \citep{Baldry_2006, Puchwein_2008, Peng_2010, Woo_2012, Chon_2016, Wang_2018, goubert2024roleenvironmentagnfeedback}. AGN feedback is the primary mechanism for galaxy quenching in most of the hydrodynamic simulations including Evolution and Assembly of GaLaxies and their Environments \citep[EAGLE,][]{Schaye_2014}, \simba \citep{Dav__2019} and IllustrisTNG \citep{Vogelsberger_2014, Piotrowska_2021}. Several papers compare feedback implementations across simulation frameworks \citep[a few are:][]{Planelles_2013, Grayson_2023,scharré2024effectsstellaragnfeedback, yang2024feedbackdrivenanisotropycircumgalacticmedium, saha2024quantifyingimpactagnfeedback, gebhardt2026cosmologicalbackreactionbaryonsdark}; however, the plethora of different modeling choices within a given simulation makes it challenging to identify the source of the large-scale structure differences. Within the same \simba simulation code, analyses have varied the specific `modes' of AGN feedback (with or without jets entirely or including jets but without X-ray emission) \citep{ sorini2022MNRAS.516..883S, Grayson_2023, scharré2024effectsstellaragnfeedback, yang2024feedbackdrivenanisotropycircumgalacticmedium}.
These comparisons showed that jet-mode feedback had the largest impact on the \textit{redistribution and heating of baryons into the WHIM phase}, as compared to X-ray and wind heating.

In this work, we extend investigations of AGN feedback by further exploring the parameter space of jet feedback prescriptions. In particular, by varying the strength of AGN feedback, we can move beyond the simple ``on/off'' test and instead probe a more continuous response of gas properties to different levels of energy injection. In doing so, we can identify thresholds and scales where feedback becomes relevant in shaping properties to match observables. Similar previous analyses within the \simba suite have varied AGN jet parameters including momentum flux, jet speed, and BH seeding \citep{Angl_s_Alc_zar_2016,pirecki2026exploringimpactagnfeedback,Grayson_2026}. 

Given the computational expense of modern hydrodynamical simulations, we choose to investigate several variations in a few zoom-in simulations. We opt to use a cluster-selected sample for resimulation, as cluster regions contain rich observational signatures that are relatively easily observed compared to lower-mass regimes. We work within the configuration of \thethr (hereafter \theth) project, a set of hydrodynamic zoom re-simulations of clusters with masses $M_{200} \gtrsim10^{14.8}\,h^{-1}\mathrm{M_\odot}$ at $z=0$ \citep{Cui_2018, Cui_2022}. The simulation allows us to investigate the impacts of gas feedback on the properties of massive clusters (at low-$z$) and in protoclusters (at higher-$z$). With a simulation high-resolution region radius of $15~h^{-1}\mathrm{cMpc}$, each region in \theth also includes filaments feeding into the clusters, plus many smaller groups. Overall, the regions contain a diverse combination of observables in a relatively small volume, making them ideal for running multiple simulations with varied physics.

In this work, we investigate how the prescriptive parameters of jets in \theth \simbac simulations affect galaxy cluster environments. The remainder of this paper is organized as follows. In Section~\ref{sec:methods}, we detail the simulation methodology, including the implementation of AGN feedback in \simbac and the parameter variations explored in our re-simulations. In Section~\ref{sec:postprocessing} we explain the post-processing tools we use to examine our data. Section~\ref{sec:results} presents the results on the effects of jet velocity caps, decoupling times, and jet orientation on the gas properties around clusters and the implications of our findings for understanding AGN feedback and large-scale structure evolution. Finally, Section \ref{sec:conclusion} summarizes our conclusions and outlines directions for future work.
\vspace{-0.1in}

\section{Simulations and setup}\label{sec:methods}
\subsection {Overview of the simulations}\label{ssec:data}

The simulations used in this work are part of \theth Project, a set of zoom-in simulations with radius $15~h^{-1}\,\mathrm{cMpc}$ centered on massive clusters from the dark-matter only MultiDark Planck 2 simulation \citep[MDPL2,][]{Klypin_2016}. 
MDPL2 has a box length of $1\,h^{-1} \mathrm{Gpc}$ containing $3840^3$ dark matter particles. The simulation assumes a fiducial \textit{Planck} cosmology with Hubble constant $H_0 = 67.77 \,\mathrm{km}s^{-1}\mathrm{Mpc}^{-1}$, cosmological matter and baryon densities of $\Omega_m = 0.307, \Omega_b = 0.048,$ clustering amplitude $\sigma_8 = 0.823$ and scalar spectral index of $n_s = 0.96$ \citep{planck2016}. After an initial simulation, clusters in the dark-matter-only MDPL2 are selected for zoom-in re-simulation on the criterion $M_{\mathrm{Vir}} > 8 \cdot 10^{14} h^{-1} M_{\odot}$ at $z = 0.$ As part of \theth project, 324 clusters are identified and then re-simulated including baryonic physics to a radius of $15~h^{-1}\,\mathrm{cMpc}$. Inside this region, particles are resolved up to a mass of $m_{\mathrm{DM}} = 1.5 \cdot 10^9\,h^{-1}\,M_{\odot}$ for dark matter and $m_{\mathrm{gas}} = 2.36 \cdot 10^8 h^{-1}\, M_{\odot}$ for gas respectively. This resolution is lower than needed to fully study galaxy properties (see \cite{G_mez_2025} for details), however our analysis is primarily focused on large-scale morphology and cluster properties rather than local galaxy properties, hence we expect resolution will not be important for the majority of findings in this paper and comment where it is a larger concern.

In our cluster re-simulations we use the \simbac galaxy evolution model \citep{2024MNRAS.532..476H}, which is a cosmological galaxy formation simulation based on \gizmo, a hydrodynamics code using the mesh-free finite-mass method \citep{2005MNRAS.364.1105S, 2014ascl.soft10003H,2017arXiv171201294H, Dav__2019}. \simbac is a variant of its predecessor \simba, with the addition of the \texttt{Chem5} chemical enrichment model and modifications to black hole seeding and feedback. It is tuned to match the $z = 0$ galaxy stellar mass function, $M_{BH} - M_{*}$ relation and quenched fraction of the galaxy population.

\simbac uses the \texttt{GRACKLE3.1} \citep{Smith_2016} prescription for radiative cooling and photoionization heating of the gas. Star formation is modelled using an $H_2$ based star-formation rate (SFR), where the SFR is given by the $H_2$ density, the dynamical time and an efficiency parameter. The $H_2$ fraction is computed following the model presented in \citep{Krumholz_2011}. \simbac tracks the chemical abundance of 34 elements $H$ to $Ge$ with enrichment from Type II SNe (SNII), Type Ia SNe (SNIa) and Asymptotic Giant Branch.  See \cite{Nomoto_2006} for details on SNII, \cite{Iwamoto_1999} for SNIa and \cite{Oppenheimer_2006} for how enrichment is implemented.

\simbac's models for star formation-driven galactic winds use decoupled two-phase winds. It includes a mass loading factor as a function of stellar mass based on the Feedback in Realistic Environments (FIRE) simulation \citep{Angl_s_Alc_zar_2017}.
\simbac uses a friends-of-friends finder on stars and gas to compute stellar mass and circular velocities \citep[see][for details]{Dav__2016}, and includes a model where BHs are seeded and fed with a two-mode accretion model, and release X-ray, wind and jet feedback. This is discussed in more detail in Section~ \ref{ssec:agn_simba}. 


\subsection{Selecting regions for zoom re-simulation}
Re-simulating the entire sample of 324 clusters with different jet properties was not computationally feasible as it takes $\sim5000$ CPU hours to run to redshift $z = 0$ (at finer mass resolutions the runtime could go up by a factor of $\sim10$). To select regions for re-simulation, we chose to focus on those which have highly anisotropic structure at $z=1$, identified through superclustering. Using these highly-clustered regions enables an investigation of the impacts of AGN feedback on the morphology of proto-clusters and their filamentary surroundings. Superclustering is often characterized by elevated overdensity and ellipticity on large scales in the dark matter or galaxy field. To choose highly-superclustered regions, we compute the projected overdensity of galaxies over the $z$ axis, 
\begin{equation}\delta_g^\mathrm{2D} = n_g^\mathrm{2D}/\bar{n_g}^\mathrm{2D} - 1,\end{equation} and smooth with a Gaussian of FWHM=10~\hMpc. We adopt the notation of \cite{1987MNRAS.226..655B} in defining dimensionless eigenvalues $\lambda_1, \lambda_2$ of the Hessian matrix, with $\lambda$ defined as positive on a peak and ordered as $\lambda_1>\lambda_2$, then calculate ellipticity as:
\begin{equation} \label{eq:ellipticity}
    e = \frac{\lambda_1 - \lambda_2}{2(\lambda_1 + \lambda_2)}.
\end{equation}
Of the 324 regions considered, we first identified those exhibiting the highest superclustering $s$ at redshift $z=1:$
\begin{equation}
    s \equiv \frac{e}{\langle e\rangle} + \frac{n_g}{\langle n_g \rangle},
\end{equation} as defined in \citet{Lokken_2023}.
Of the highest-ranked 15 regions, we then select three that contain high numbers of BHs and average BH accretion rate within the simulation volume. 
We emphasize that due to this selection, this work analyzes impacts of AGN in specialized regions of the cosmic web with elevated overdensity and high ellipticity, that may not be representative of the average. We leave analysis of all 324 cluster regions to future work.

\subsection{AGN feedback in \simbac\label{ssec:agn_simba}}

In the \simbac model, BHs can be seeded into galaxies with masses as low as $6 \cdot 10^8 M_{\odot}$, but their feedback is suppressed by an exponential factor $\text{exp}(-M_{BH}/(10^6 M_{\odot}))$. In previous \simba models, BHs were seeded into any galaxy with stellar mass cutoff $M_\star \gtrsim 5 \cdot 10^9 M_{\odot}$, if it did not already have a BH. The exponential suppression of BH growth was included to match black hole growth seen in zoom simulations in which star formation feedback suppressed gas supply \citep{Angl_s_Alc_zar_2017, Hopkins_2021}. The approach allows us to track BH growth at much higher redshifts (see \citealt{Hough_2023} for a more detailed discussion on the BH seeding mechanism in \simbac). This alteration is well suited to our needs, as we are interested in probing specific redshifts when AGN feedback effects dominate. AGN feedback is implemented into the \simbac simulation through the radiative and jet channels (\citealp{Angl_s_Alc_zar_2017, Hopkins_2021,Cui_2022}; Cui et al., in prep.) :
\begin{eqnarray}
  v_{w,\mathrm{Radiative}} &=& 500 + \frac{500}{3}\bigg(\log_{10}\frac{M_{\mathrm{BH}}}{M_{\odot}}-6 \bigg) \text{ km s}^{-1}\\
v_{w,\mathrm{Jet}} &=& v_{w,\mathrm{Radiative}} + v_{\mathrm{Max}} \log_{10} \bigg(\frac{0.2}{f_{\mathrm{Edd}}}\bigg) \text{ km s}^{-1}\nonumber \\
&& \quad\quad\quad\quad\quad\quad\quad 0.02 < f_{\mathrm{Edd}}<0.2. \label{eq:jet_vel}
\end{eqnarray}

Here $M_{\mathrm{BH}}$ is the central BH mass, $f_{\mathrm{Edd}}$ is the Eddington ratio and $v_{\mathrm{Max}}$ is a jet velocity cap. The Eddington ratio is given by $f_\mathrm{Edd} = \frac{\dot{M}_\mathrm{BH}}{\dot{M}_\mathrm{Edd}}$, where $\dot{M}_\mathrm{Edd}$ is the BH's Eddington accretion rate. The jet mode is only activated when $f_{\text{Edd}}<0.2.$ In addition, a cap on the velocity of the jet boost to prevent it from exceeding $v_{\text{Max}}$. Thus, BHs with Eddington ratios below $f_{\text{Edd}}<0.02$ will all be boosted by the same amount. To trigger the jet mode the minimum BH mass has to be greater than $10^{7.5} M_\odot$. Here the feedback modes are controlled by the Eddington ratio $f_{\text{Edd}}$. An X-ray feedback mode following \cite{Choi_2012} is also incorporated when the jet conditions are met, plus the additional criteria that $M_* > 10^9 \mathrm{M_{\odot}}$ and gas fraction $m_{\mathrm{gas}}/m_{\mathrm{baryon}} < 0.2$.

\begin{figure}[htbp!]
    \centering
\includegraphics[width=0.85\linewidth]{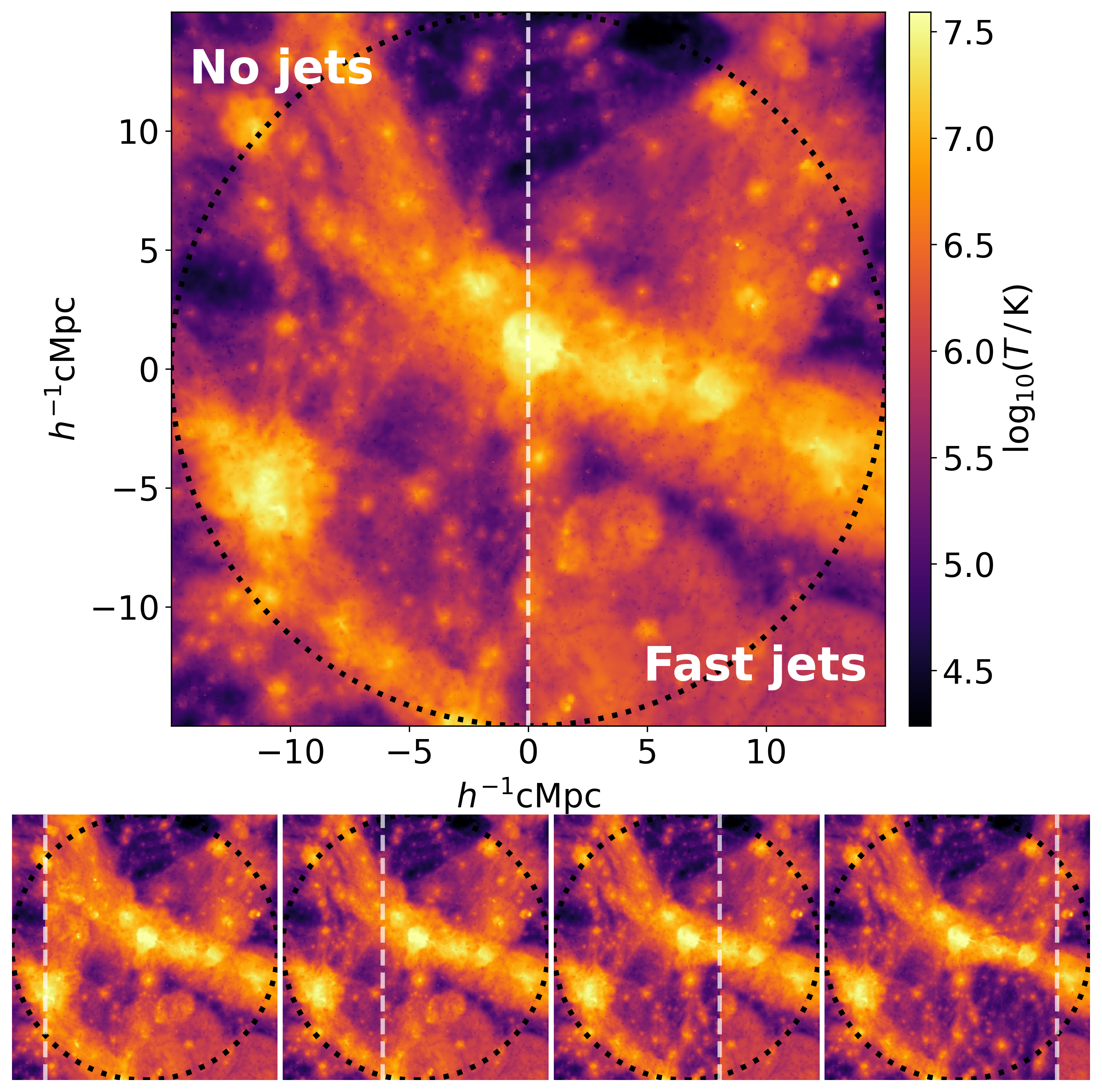}
    \caption{A mass-weighted gas temperature projection of the no jets and fast jets runs, side by side at redshift $z = 1$. The total projection is a cube with a side length of $30~h^{-1} \,\mathrm{cMpc} $. In the upper figure, the left-hand side shows the no jet run while the right-hand side shows the fast jet run, with the dotted white line denoting the transition between simulations. Both halves follow the same coloring. The bottom row of plots are the same as above, but show the transition in different locations. The black dotted circle denotes the trustworthy region of the simulation, which has a radius of $15~ h^{-1} \mathrm{cMpc}$. To make the projection clearer, we omit particles from dense star-forming regions, with star formation rates above $0.1 M_{\odot}/\mathrm{yr}$ and densities above $3 \cdot10^{10}h^2M_{\odot}/\mathrm{ckpc^3}$. }
    \label{fig:half_half}
\end{figure}
We define the BH accretion rate $\dot{M}_\mathrm{BH}$ as the sum of the Bondi and torque-driven modes, as introduced in \cite{Angl_s_Alc_zar_2016} and implemented in \simba and \simbac \citep[][respectively]{Dav__2019, Hough_2023}. We use the torque-limited accretion for cold gas ($T < 10^5~\mathrm{K}$) and Bondi-Hoyle-Lyttleton  accretion for hot gas \citep{bondi/hoyle:1944, hoyle/lyttleton:1939, 1952MNRAS.112..195B}:
\begin{eqnarray}
\dot{M}_{\mathrm{Torque}} &\equiv\; & 
\epsilon_{T} f_{d}^{5/2}
\left( \frac{M_{\mathrm{BH}}}{10^{8} \, M_{\odot}} \right)^{1/6}
\left( \frac{M_{\mathrm{enc}}(R_{0})}{10^{9} \, M_{\odot}} \right) \nonumber  \\
&& \times 
\left( \frac{R_{0}}{100 \, \mathrm{pc}} \right)^{-3/2}
\left( 1 + \frac{f_{0}}{f_{\mathrm{gas}}} \right)^{-1}
\, M_{\odot} \, \mathrm{yr}^{-1},
\end{eqnarray}
\begin{equation}
\dot{M}_{\mathrm{Bondi}} = \frac{4 \pi G^2 M_{\mathrm{BH}}^2 \rho}{c_s^3}.
\end{equation}
Here $\dot{M}_{\mathrm{Torque}}$ is following the definition in \cite{Hopkins_2012}, and the Bondi accretion from \cite{1952MNRAS.112..195B}. See \cite{Dav__2019} for more details on BH growth and accretion within \simba. The total accretion rate for a given BH combines these two modes:
\begin{equation}
    \dot{M}_{\mathrm{BH}}= (1 - \eta) \times (\dot{M}_{\mathrm{Torque}} + \dot{M}_{\mathrm{Bondi}})
\end{equation}
where $\eta = 0.1$ is the radiative efficiency \citep{Yu_2002}. The conditions for the jet velocity kicks in Equation~\ref{eq:jet_vel} are motivated by observational studies where $f_{Edd}$ determines the AGN feedback mode. They suggest that at high Eddington ratios the AGN operates in a ``radio-quiet'' mode, depositing energy thermally into the surrounding medium, while at lower Eddington ratios the AGN enters a ``radio-loud'' mode, where energy is injected kinetically via collimated jets (see e.g. \citealt{2002ApJ...564..120H, 2007ApJ...658..815S}, although we note that some e.g. \citealt{Ballo_2012, long2025revisitingfundamentalplanesblack} have questioned the role of Eddington ratio in AGN feedback). Both the winds and the jets are mass-loaded via: \begin{equation}
    \dot{P}_{\text{out}} = 20\, c \eta \dot{M}_\text{BH},\label{eq:mass_load}
\end{equation}
where $c$ is the speed of light, $\eta = 0.1$ and $\dot{M}_{\text{BH}}$ is the BH accretion rate. It is believed that real jets are ejected at relativistic speeds and mass-loaded. Incorporating such velocities into the simulation would lead to numerical instabilities from high mach shocks in very dense gas near the BH. In addition, resolution limits of the simulation would lead to artificial interactions with the ISM that would not happen if we could properly resolve the jet. Our prescription is intended to represent the energy and momentum once the jet reaches kpc scales.  To account for the highly collimated nature of jets that is also unresolvable, \simbac sets a time after the jet is ejected for which it travels freely, decoupled from the surrounding media: \begin{equation}
    t_{\text{Decoupling}} = \alpha \cdot t_{\text{Hubble}},\label{eq:t_decoup}
\end{equation}
where $\alpha$ is a scaling constant. This also prevents the jets from interacting with ISM near the central BH, which is not needed as mass-loading is already given to the jets by Equation~\ref{eq:mass_load}, as shown in \citet{Dav__2019}. 

\subsection{Parameter variations}\label{ssec:param_runs}
We configure six sets of AGN feedback parameters, shown in Table~\ref{tab:runs_summary} and illustrated in Figure~\ref{fig:jet_props}, which we run across simulation regions. We vary the maximum jet velocity, emission direction, and decoupling time. In the \simba simulations, the maximum velocity was capped at $v_{\mathrm{Max}}=7,000~\text{km s}^{-1}$, while in \theth \gizmo-\simba run the cap was changed to $v_{\mathrm{Max}}=15,000~\text{km s}^{-1}$. This value is somewhat arbitrary, and has been set to loosely calibrate to radio and cluster observations. Hence we explore the impacts of changing the jet velocity cap. On the other hand, if the jet speed were lowered it would simply mimic the effects of winds; therefore as another point of comparison we turn jets off entirely \citep[a variation that was also explored in the original \simba simulations, see][]{Dav__2019}. We perform two variations to the baseline: one where $v_{\mathrm{Max}}=35,000~\text{km s}^{-1}$ and another where jet feedback is disabled entirely. Beyond $35,000~\text{km s}^{-1}$, which is already over $10\%$ of the speed of light, some numerical instabilities may begin to occur in the modelling. As a result, our choices for the jet velocities are designed to compare a calibrated model with one near its upper limits.

Figure~\ref{fig:half_half} shows the difference in projected gas temperature across a region for the no jet and fast jet runs. The mass-loading remains constant for all of our runs, so in changing the jet velocity we are also modifying the momentum flux.

Under previous \simba runs, the decoupling time is defined with $\alpha = 10^{-4}$, which allows the highest velocity jets to travel approximately 
$10~\mathrm{kpc}$ before recoupling with the surrounding gas, which is around the radius of a galaxy. We perform an additional run with $\alpha = 2 \cdot 10^{-3},$ where jets can travel up to hundreds of \text{comoving} \text{kpc} before interaction. We do not include runs with shorter decoupling time for reasons discussed in Section~\ref{ssec:agn_simba}.

The third parameter which we vary is the jet emission direction. There are two main pictures of gas inflow to supermassive black holes (SMBHs): the coherent and chaotic models. Under the coherent model, SMBHs are smoothly spun up by the steady accretion of gas, leading to AGN accretion perpendicular to the plane of the galaxy \citep{2011MNRAS.414.2148L}. This aligns with recent observational work, which has found that radio AGN jets appear preferentially aligned with the minor axis of the host galaxy \citep{Zheng_2024}. The chaotic model suggests that there are major shifts in the angular momentum axis during the accretion process; the resulting AGN jets are randomly aligned relative to their host galaxy. We consider two models of AGN jet directions to represent these coherent and chaotic scenarios. The first where feedback is emitted in purely bipolar outflows, with direction dictated by the angular momentum of the gas particles in the BH kernel, which has a radius containing the 256 nearest gas particles, or all the particles within $6\ h^{-1}\mathrm{ckpc}$, whichever is smaller. In the other model the jet direction is randomized for each galaxy and timestep. The bi-polar jets have been a feature of \simbac and \simba, whereas the random emission direction is used in the IllustrisTNG simulations \citep{Weinberger_2016, Marinacci_2018}.

\vspace{-0.1in}
\section{Post-processing and analysis tools}\label{sec:postprocessing}
After running the simulations, for each run we are left with 129 snapshots from redshift $z=17$ to $z = 0$. While some fields like the particle positions and masses are stored natively (see \citealt{Hopkins_2012} for details), quantities relevant to AGN feedback such as halo membership, thermal observables, and large-scale structure must be derived in post-processing. The subsections below describe the tools and methods we use to extract these quantities, including halo identification, mock tSZ map generation, overdensity classification, and cosmic filament detection.

\begin{figure*}[htbp!]
    \centering
\includegraphics[width=0.65\linewidth]{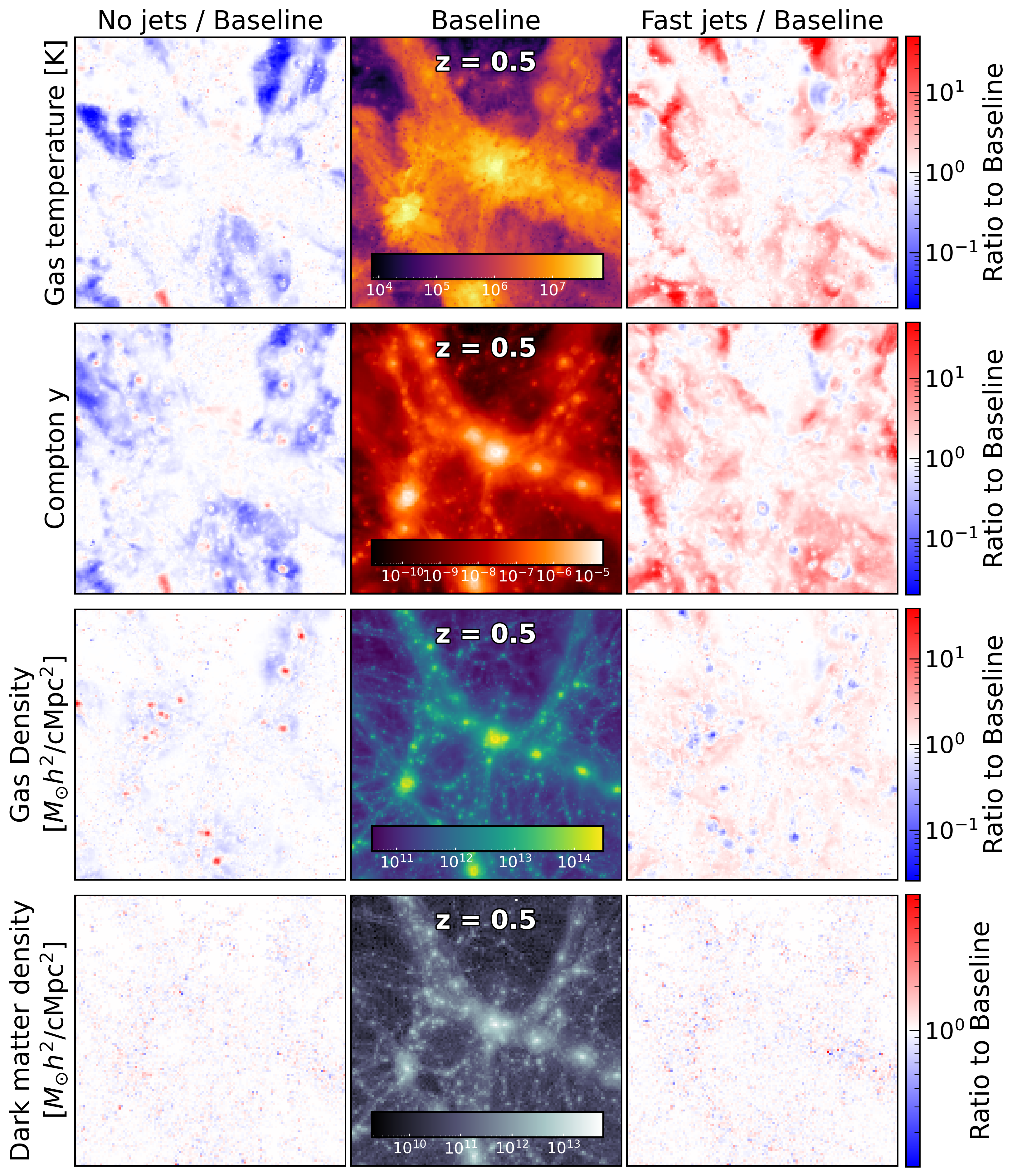}
    \caption{In each box of the plot above is the projection of a box with dimensions $30\, h^{-1}\mathrm{cMpc} \times 30\,h^{-1} \mathrm{cMpc} \times 10\, h^{-1}\mathrm{cMpc}$ at redshift $z = 0.5$. The columns are labeled according to the parameter runs and variation in the jet model (``No jets'', ``Baseline'' and ``Fast jet''), and the left- and right- columns are shown as increases or decreases relative to the baseline model shown in the middle column (shown with a common color calibration per row).
    Each row shows a different tracer: the mass-weighted gas temperature (top), Compton $y$-map (second), gas density (third) and dark matter density (bottom). We note that the differences in dark matter density between the different jet models is negligible as expected, however large differences show up in the gas temperature and the Compton $y$ signal.}
    \label{fig:jet_vel_tracers}
\end{figure*}

\subsection{Halo-finding}
Halos are identified using a 3D friends-of-friends algorithm built in \gizmo, based on the code of GADGET-3. After running the \simba simulation, we use the YT-based package \caesar\footnote{https://caesar.readthedocs.io/en/latest/} to cross-match galaxies and halos for every snapshot. \caesar also produces catalogues of several halo properties that are used throughout our analysis, including $ R_{500}$ and the halo stellar, gas and BH masses.

\begin{figure*}[htbp!]
    \centering
    \includegraphics[width=0.75\linewidth]{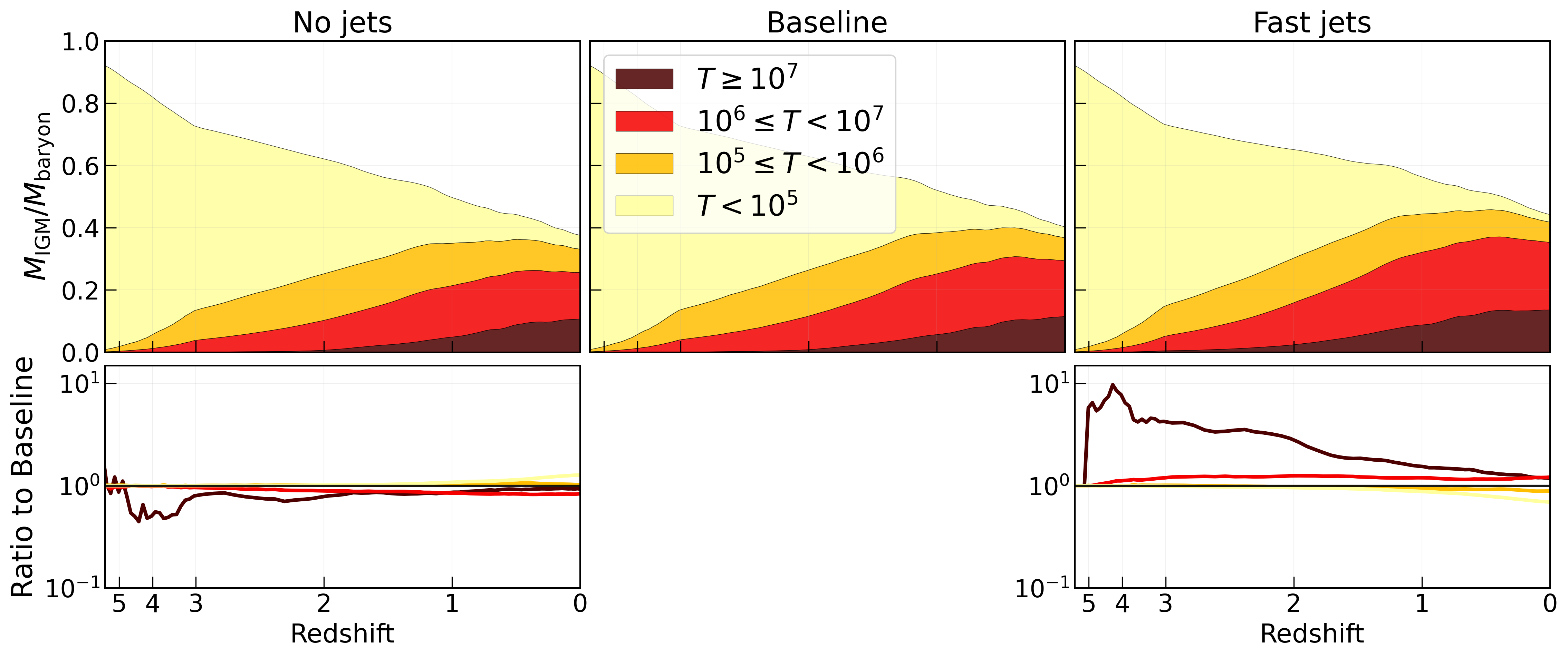}
        \caption{The mass-weighted fractions of IGM gas for three different jet velocity runs on region A as a function of redshift. IGM gas is identified as any gas particles not associated with dark matter halos, within $15 \, h^{-1} \mathrm{cMpc}$ from the central cluster. The IGM gas fractions is defined as the mass of IGM gas in various temperature bins divided by the total baryonic mass in the simulation volume. The bottom panel illustrates the fraction relative to the baseline.
    \label{fig:igm_temp}}
\end{figure*}

\subsection{Thermal Sunyaev-Zel'dovich effect} \label{subsec:tSZmaps}



Feedback influences the temperature and density of gas beyond the AGN hosts, hence the tSZ effect probes these changes. The tSZ effect arises from inverse Compton scattering, where cosmic microwave background photons gain energy by interacting with high-energy electrons \citep{sunyaev1970small,sunyaev1972observations}, and produces a distortion in the CMB frequency spectrum in sightlines containing hot ionized gas. In simulations, we can compute mock tSZ maps to compare with observations by integrating the electron pressure along the line of sight. The Compton $y$-parameter is defined as:
\begin{equation} y =  \int \frac{k_B \sigma_T}{m_e c^2} n_e T_e dl,\end{equation} 
where $k_B$ is Boltzmann’s constant, $m_e$ is the electron mass, $c$ is the speed of light, and $n_e$ and $T_e$ are the electron density and temperature.
We create mock Compton $y$-maps using the \pymsz \footnote{https://github.com/weiguangcui/pymsz} package \citep{Cui_2018}. This uses the SZPACK library, which allows for fast computations of the tSZ signal \citep{Chluba_2012, Chluba_2013}. Each mock $y$-map is projected over one axis of the simulation box. Figure~\ref{fig:jet_vel_tracers} shows a comparison of the gas temperature, tSZ signal, gas density and dark matter density for the jet velocity variations. The gas temperature is the most strongly impacted by the jet speed variations, which propagates to impacts on the $y$ signal. The effect on the gas density is more subtle; dense gas in small halos becomes more dispersed with higher jet velocities. The effects on the dark matter density are negligible in comparison. To make the projection clearer, we omit dense star-forming regions, with star formation rates above $0.1 \, M_{\odot}/\mathrm{yr}$ and densities above $3 \cdot10^{10}\,h^2M_{\odot}/\mathrm{ckpc^3}$. Without these cuts, these dense areas would appear as dark dots in our projections as an artifact of weighing the gas temperature by a very high density. This cut was chosen qualitatively, and removes $\sim10 \%$ of the gas mass.

The $y$-maps are generated with an angular resolution of 0.5 arcminutes per pixel (at redshifts $z=0.5,1$). We then apply a Gaussian kernel smoothing with a FWHM of $\theta_{\mathrm{FWHM}} = 1.4^\prime $. This was chosen to match the resolution of the of the Simons Observatory (SO) Large Aperture Telescope in the 145~GHz band \citep{thesimonsobservatorycollaboration2025simonsobservatorysciencegoals}. At redshift of $z=0$, the $y$-map is generated with a fixed resolution of $500\times500$ pixels without additional smoothing.
\subsection{Overdensity}\label{ssec:overdensity}

We define gas overdensity as the deviation of the local and global densities as:\begin{eqnarray} 
\bar{\rho} &=& \Omega_b\rho_c; \ \ \rho_c = \frac{3 H(z)^2}{8 \pi G}\label{eq:critical_density}
\\
        \delta ({x}) &=& \rho({x}) / \bar{\rho} - 1, \label{eq:overdensity}
\end{eqnarray}
 where $\rho$ is the density at $x$, $H(z)$ is the Hubble parameter and $\bar{\rho}$ is the critical mass density multiplied by the baryon mass density parameter. For simplicity, we compute the overdensity using only the gas particles, and not including star or dark matter particles. As shown in the lower two rows of Figure \ref{fig:jet_vel_tracers}, the gas and dark matter densities are largely aligned with each other. We leave analysis using the dark matter overdensity field for future work. The left panel of Figure \ref{fig:gas_frac_z} demonstrates that the gas overdensity field effectively traces the structure of the cosmic web. \cite{Arag_n_Calvo_2010} show that cosmic voids, filaments, walls and nodes have distinct overdensity distributions, but overdensity cannot be used alone for classification. Nevertheless, we use it as a simple diagnostic for cosmic environments. 

 \subsection{Filament-finding algorithms}\label{ssec:dispere}
We  use the Discrete Persistent Structure Extractor \citep[\textsc{DisPerSE}][]{Sousbie_2011} to identify filamentary structures in our simulations. This method is based on discrete Morse theory \citep{FORMAN199890, gyulassy2008combinatorial} and detects topological features such as critical points, filaments, walls, and voids, which are associated with different components of the cosmic web over a wide range of scales. \textsc{DisPerSE} returns the spatial coordinates of critical points, or locations where the gradient of the underlying field is null. Filaments are constructed by connecting pairs of critical points along integral lines of the gradient field. The statistical significance of each filament is characterized by its persistence, a scalar quantity that measures the contrast between the field values at the two critical points linked by the filament. By imposing a minimum persistence threshold, structures dominated by noise or low-contrast fluctuations are excluded, leaving robust filamentary features. To identify cosmic filaments, we apply the \textsc{DiSperSE} algorithm to the mock $y$-maps described in Sec.~\ref{subsec:tSZmaps}. On real data, one would typically apply a filament-finding algorithm on galaxies because $y$-maps are too noisy to identify these low-SNR structures. However, we take advantage of our noiseless simulated maps to identify the pressure filaments directly from the $y$-map.

To identify the filament skeleton, we run \textsc{DiSperSE} on the baseline simulation only, then assume the filaments have the same exact locations in the other feedback runs. This avoids noise in the comparison. We only perform this analysis using simulation region A, and one projection direction. Prior to running \textsc{DiSperSE}, the baseline $y$-map is smoothed with a Gaussian kernel ($\mathrm{FWHM} = 0.9 \ \mathrm{kpc}$) and converted into $\log_{10}(y)$, to improve the ability of \textsc{DiSperSE} to identify large-scale filaments. Lastly, we apply \textsc{DiSperSE} with a persistence absolute cut (the difference between the density values of two critical point pair) of 1 and 1.5, respectively at redshift $z=1$ and $z=0.5$. Here, the persistence absolute cut is in the same units as the maps \citep{Santoni_2024, santoni2026projectcosmicwebidentification}.

\section{Results and Discussion}\label{sec:results}We quantitatively analyze the impacts of varying jet velocity on the gas temperature, density, morphology, as well as several galaxy and BH properties. We find that the impacts from varying jet direction and decoupling time are weaker, so the sections below focus on comparing the results from different jet velocities.

\begin{figure*}[htbp!]
\centering
\includegraphics[width=0.3\linewidth]{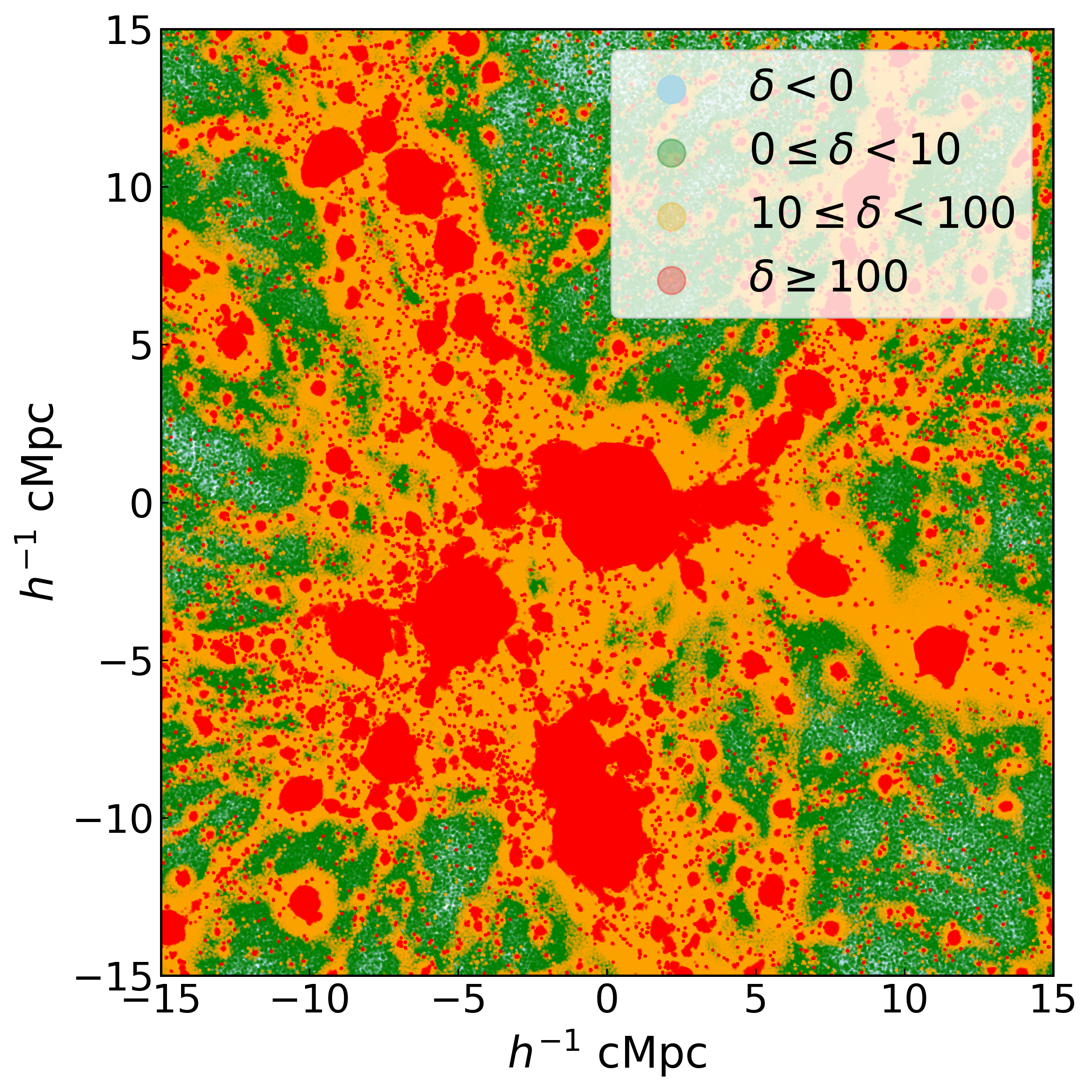} 
\includegraphics[width=0.65\linewidth]{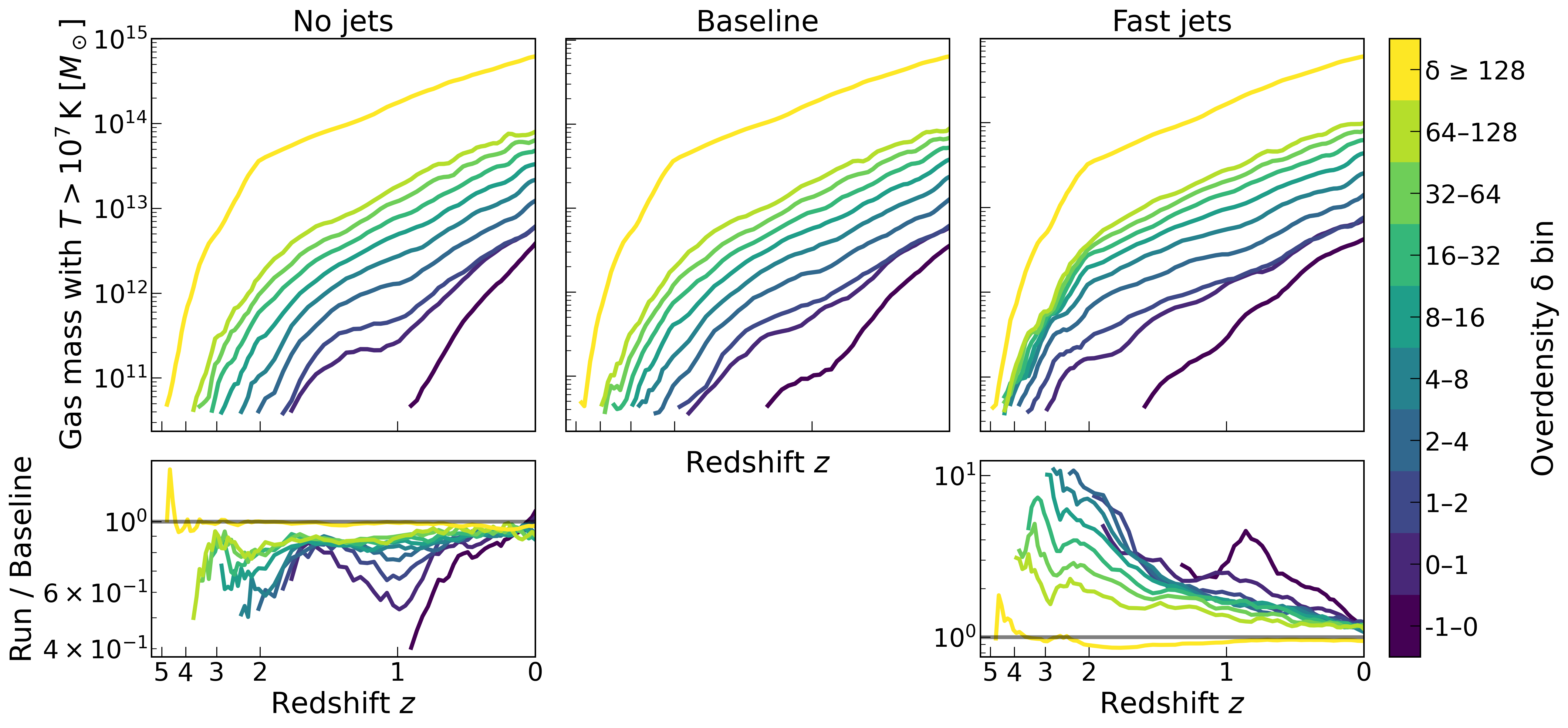}
        \caption{Left: a $30~h^{-1}\mathrm{cMpc}\times 30~h^{-1}\mathrm{cMpc}\times 30~h^{-1}\mathrm{cMpc}$ projection of the overdensity field at redshift $z=0$. The particles are colored by their value of overdensity in bins: $\delta > 100$ (red), $10 < \delta < 100$ (yellow), $0 < \delta < 10$ (green) and $\delta < 0$ (blue). These overdensities are used to illustrate trends in the behaviour of the gas at different temperatures as shown on right and Figure~\ref{fig:decoupling_overdensity}. Right: the total gas mass in $M_{\odot}$ above $T>10^7\, \mathrm{K}$, divided by separate lines for overdensity, within $15\  h^{-1}  \mathrm{cMpc}$. The three plots in the top row show the mass for three different jet velocity runs of region A, while the bottom panel illustrates the fraction relative to the baseline. }\label{fig:gas_frac_z} 
\end{figure*}

\subsection{Temperature evolution in IGM}
We examine the properties of baryons in the IGM by running \caesar and labeling any gas particle not associated with a dark matter halo as IGM. In Figure \ref{fig:igm_temp} we compute the fraction of IGM gas in various temperature bins as a function of redshift. This is similar to the comparisons of IGM gas across the \simba simulations computed in \citep[][hereafter S22]{sorini2022MNRAS.516..883S}. Their Figure 2 showed that gas in the WHIM was most significantly impacted by inclusion of AGN jets. A key difference is that, because we center on clusters, our simulations contain a larger fraction of hot gas ($T>10^7$~K) than \simba. In addition, towards lower redshift, most of the gas has left the IGM and collapsed onto halos. At high redshifts we find that the fraction of hot gas is significantly impacted by the jet properties; the overall fraction is small in all cases, but fast-jet heating amplifies it by an order of magnitude. However, we find that this component is not significantly impacted by the jet variations by $z=0$. In S22 the fraction of warm-hot gas at $z=0$ ($10^6\,\mathrm{K}<T<10^7\,\mathrm{K}$) reduces by more than a factor of two when jets are turned off, but in our simulations it only varies by $\sim20\%$ across the runs. This is likely due to the specialized regions we simulate, which are centered on clusters, as opposed to S22 which uses a $50~h^{-1}\,\mathrm{cMpc}$ box.

\subsection{Temperature evolution across overdensity regimes}\label{ssec:overdensity}

In Figure \ref{fig:gas_frac_z} we examine the gas mass above $10^7\, \mathrm{K}$ at different gas overdensity scales, where the bottom row shows the ratio as compared with the baseline run. A minimum of 100 gas particles must be found in a given density bin to be included. We see that in denser regions, shown by the lighter colored lines, the temperatures are largely consistent across the three different runs, suggesting that feedback does not heavily impact gas temperature in clusters and filaments.
In the most diffuse regions, represented by the darker blue lines, we see that jet velocity has a very strong effect on temperature. In these regions, corresponding to overdensities below 8, the fast jet run has up to nearly an order of magnitude more gas above $10^7\,\mathrm{K}$ around $z \sim 1 -4$. In other words, strong AGN jet feedback is needed to push any hot gas into diffuse regions at higher redshifts. Conversely, there is approximately two times less hot gas in the most diffuse regions in the no jet run. We note that we are again limited by simulation volume where low density areas are near massive clusters. As a result, these are the voids that would be most susceptible to jet-induced heating. Regions that are more isolated from dense clusters would likely remain cooler. Due to these selection effects, there are very few gas particles in underdense regions within the simulation volume, let alone underdense particles above $10^7 \ \mathrm{K}$. 

\begin{table}[]
    \centering
    \begin{tabular}{c|c|c|c}
        Run & Region & $\langle y\;(R>5\;\mathrm{Mpc}) \rangle$ & $\langle y\rangle$ \\
        No jets & A & $8.0\times10^{-8}$ & $1.4\times10^{-7}$\\
        Baseline &  A & $8.6\times10^{-8}$ & $1.5\times10^{-7}$\\
        Fast jets & A & $9.5 \times10^{-8}$ & $1.6\times10^{-7}$ \\
        Random & A & $8.5\times10^{-8}$ & $1.5\times10^{-7}$\\
        No jets & C & $2.7\times10^{-8}$  &$1.0\times10^{-7}$ \\
        Baseline & C & $3.1\times10^{-8}$ & $1.1\times10^{-7}$ \\
        Fast jets & C & $3.7 \times10^{-8}$ & $1.3\times10^{-7}$ \\
        Late coupling & C & $3.1\times10^{-8}$ & $1.2\times10^{-7}$\\
    \end{tabular}
    \caption{The mean-$y$ measured for three axes of each simulation, both outside the central cluster (third column) and across the full map. The maximum radius included is $R=20$~Mpc to avoid edge effects. Higher jet velocities increase the mean-$y$, having a larger fractional impact beyond the central cluster.}
    \label{tab:mean_y}
\end{table}

The denser regions, represented by the lighter lines, are less affected by the changes in feedback model. This is not surprising as previous \theth analysis has shown that even with different baryonic models (\gadgetx and \simba) the hot gas mass within halos at $z = 0$ only varies by $\sim 2\%$ \citep{Li_2023}.

We see very large effects in the hot gas mass around redshift $z=4$, which, as shown in Figure \ref{fig:jet_props}, is around the time when the first jets are turning on. As a result, the large difference in hot gas mass in underdense regions comes from just one or a few jets which have heated their surroundings. At such high redshifts, many group-scale halos had not yet merged, and the filamentary structures had not yet formed. A jet leaving from a BH would not have to travel very far in order to reach what we label as an underdense region and so jets can have an outsized impact at high redshift. Lastly, we are defining the overdensity as described in Section~\ref{ssec:overdensity} using only the gas particles. Using the dark matter field could provide a clearer picture on the effects of AGN feedback-induced heating on different cosmic environments, which we leave to future work.

\vspace{-0.1in}
\subsection{Mean tSZ}

\begin{figure*}[htbp!]
    \centering
    \includegraphics[width=0.95\linewidth]{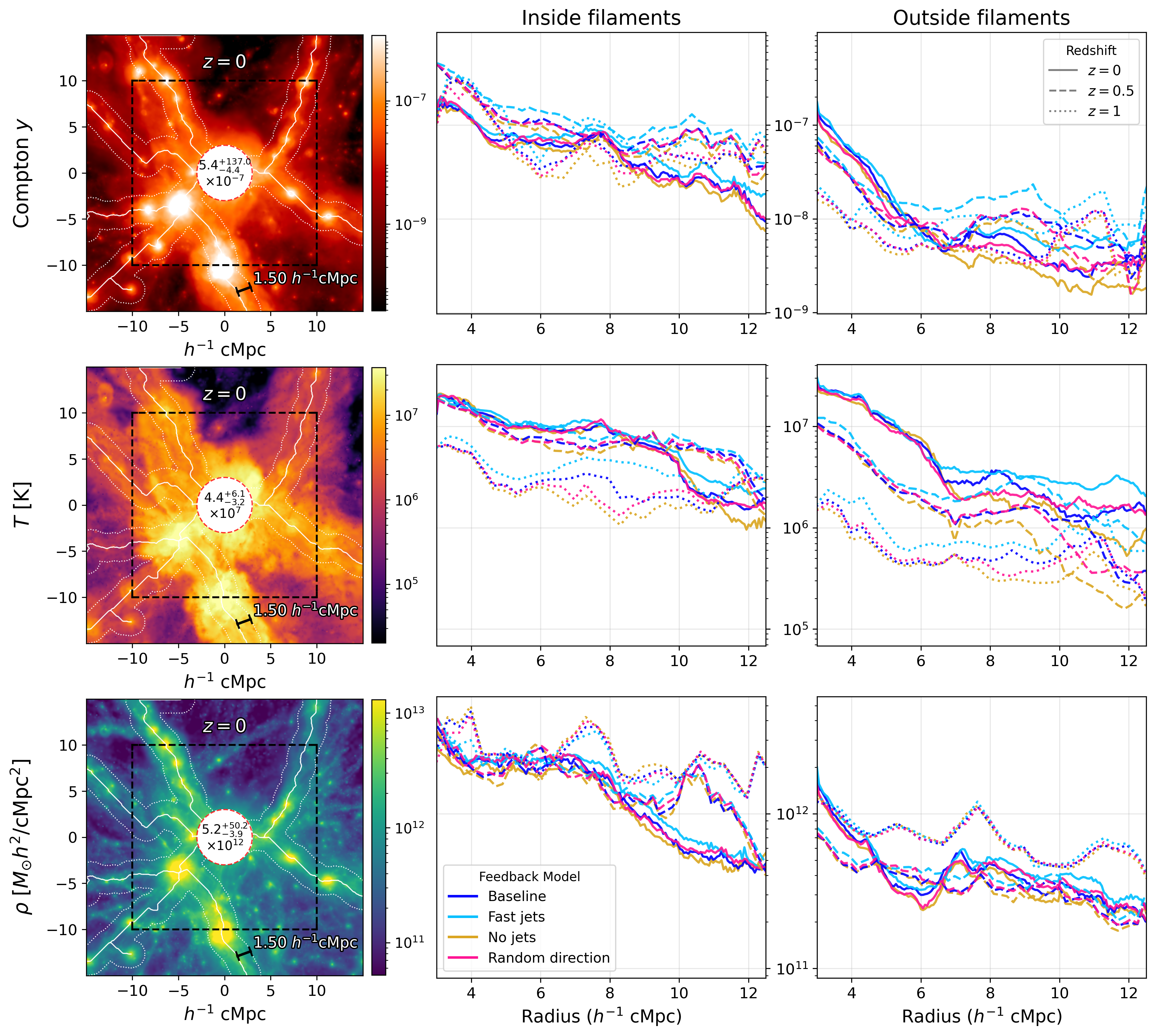}
    \caption{The center and right columns show the median Compton $y$ (top row), temperature (middle row) and gas density (bottom row) in circular shells as a function of radius from the central cluster. The left column is the Compton $y$, temperature and gas density, respectively, at redshift $z = 0$ for the baseline run. The white solid line and the white dotted lines surrounding it denote the portion of the map deemed to be filaments, which are set to width $1.5\, h^{-1}\mathrm{cMpc}$. The white circle is the central cluster, which is masked from the profile but its median and 95\% confidence interval for the baseline run is overlaid. The black dotted box represents the box used for the profiles with dimensions $[20 \, h^{-1} \mathrm{cMpc} \times 20 \, h^{-1} \mathrm{cMpc} \times 10 \, h^{-1}\mathrm{cMpc}]$, while a larger box size of $[30 \, h^{-1}\mathrm{cMpc} \times 30 \, h^{-1}\mathrm{cMpc} \times 30\,h^{-1} \mathrm{cMpc}]$ was used to calculate the filament skeleton. The middle column is the median value of the field inside of filaments as a function of radius, while the right column is outside of filaments. Each feedback model is plotted with different colours. The three line styles represent $z=0$ (solid line), $z= 0.5$ (dashed line) and $z = 1$ (dotted line).}
    \label{fig:sz_fil}
\end{figure*}

The simplest statistic to compute on the mock tSZ maps is the mean. COBE FIRAS, an early CMB instrument, aimed to measure the absolute temperature of the CMB, and thus contained sensitive spectrophotometers enabling an observational bound on the $y$ monopole \citep{Fixsen1996ApJ...473..576F}. Recent work has improved the constraint and demonstrated the usefulness of $\langle y \rangle$ for constraining feedback \citep{Fabbian2025arXiv251203038F}. However, for instruments like ACT, \textit{Planck}, and SO that have been designed to measure only the CMB temperature distortions, rather than absolute value, the mean-$y$ information is extremely weak and becomes lost in the process of linear map combination. Proposed CMB spectral distortion experiments like BISOU \citep{Maffei2024SPIE13102E..0NM} and FOSSIL \citep{Aghanim_FOSSIL} would be able to improve measurements of the mean-$y$.

Given that we are limited to single snapshots rather than projected lightcones, and our re-simulated regions are overdense with respect to the average universe by nature of being centered on clusters, we do not expect to measure a realistic global $\langle y \rangle$, but examining the impacts of the feedback variations on $\langle y_\mathrm{box} \rangle$ can still shed light on the expected relationship between jet parameters and $\langle y \rangle$. We measure the total mean, including the inner 5~Mpc, and the mean beyond 5~Mpc to exclude the central cluster, for each run. The results are reported in Table~\ref{tab:mean_y} for $z=0.5$. Increased jet velocities increase the mean $y$, especially impacting the lower-density regions beyond the central cluster, which is consistent with the temperature impacts in low-density regions discussed in the previous subsection. When jets are oriented randomly or are further delayed in their hydrodynamic coupling, $\langle y \rangle$ is similar to the baseline run, indicating that these variations have a smaller impact on the average temperature. For reference, \citet{Fabbian2025arXiv251203038F} predict a Fisher error on the global $\langle y \rangle$ of $\sigma_{\langle y \rangle }\sim 1.6\times10^{-8}$ for FOSSIL, similar in magnitude to the difference between no jets and fast jets in Table~\ref{tab:mean_y}, indicating this instrument might be able to discriminate between such feedback models in cluster environments.

\vspace{-0.1in}
\subsection{tSZ in filaments}
We investigate how the impact of feedback varies across different cosmic environments by separating the tSZ signal into filamentary and non-filamentary regions.

The \textsc{DiSperSE} output provides only a skeleton tracing the filament spines, without explicit information about their extent. Previous insight into the physical extent of filaments in cluster environments is provided by recent analyses of \theth simulations \citep{Kuchner_2020,rost2023projectthermodynamicalpropertiesshocks}. Using a filament-finding algorithm applied to the full three-dimensional gas distribution, \cite{Kuchner_2020, rost2023projectthermodynamicalpropertiesshocks} examine the distribution of gas particles around the spine to find characteristic filament radii of order $\sim 0.7$--$2 \,h^{-1}\mathrm{Mpc}$, with substantial variation as a function of distance from the cluster center and local dynamical state. 

In this work, we adopt a fixed filament radius of $1.5\, h^{-1} \mathrm{cMpc}$ for simplicity and defer a more detailed treatment of filament geometry and its radial and redshift evolution to future work. All pixels located within this radius from the filament spine are classified as being inside filaments, while all remaining pixels are classified as outside filaments.  In Figure~\ref{fig:sz_fil}, we show the radial Compton-$y$, temperature and gas density profiles measured separately inside and outside of filaments. The Compton-$y$ profile is computed from the baseline $y$-map with SO-like pixelization and resolution. Starting at a projected radius of $3\, h^{-1}\mathrm{cMpc}$ from the cluster center, we compute the median parameter value in concentric circular annuli extending to the edge of the map. By starting away from the center, we can more accurately compare the effects of feedback in and outside of filaments. For context, the median value and 95\% confidence interval of the field for the baseline run within the masked region is printed inside the white circle. We note that while the central cluster only has a size of $R_{500}\sim2\,\mathrm{Mpc}$, its thermal structure clearly extends far beyond that.

Both inside and outside filaments, stronger feedback results in an enhanced Compton-$y$. Within filaments, the relative differences between feedback models are of order $\sim 10\%$, whereas outside filaments they are typically at the $\sim 100\%$ level. Since the Compton-$y$ parameter traces the integrated thermal pressure of the gas, the signal drops off sharply at large distances from both the cluster center and filamentary structures. As a result, despite proposed large differences in the Compton-$y$ outside of filaments, this would be difficult to detect observationally.

By comparison with the second and third rows, looking at the temperature and gas density, respectively, we see that the differences in the Compton-$y$ are mostly driven by changes in temperature. 

The gas density profiles at $z=1$ appear a few times higher than those at lower redshift. This is an artifact of the fixed angular beam, $\theta_{\mathrm{FWHM}} = 1.4^\prime$. Because angular diameter distance increases with redshift (in this range), this fixed beam corresponds to a larger physical smoothing scale at high-$z$. When applied to the right-skewed density field, this larger kernel spreads the high-density cluster core signal into surrounding low-density pixels, raising the median.

We also see that changing the jet emission direction has very little effect on the gas temperature and Compton-$y$, as compared to the jet velocity variations. The late coupling run was only computed on simulation region C and thus excluded from this analysis.

\vspace{-0.1in}
\subsection{Oriented stacking}
\begin{figure*}
    \centering
    \includegraphics[width=\linewidth]{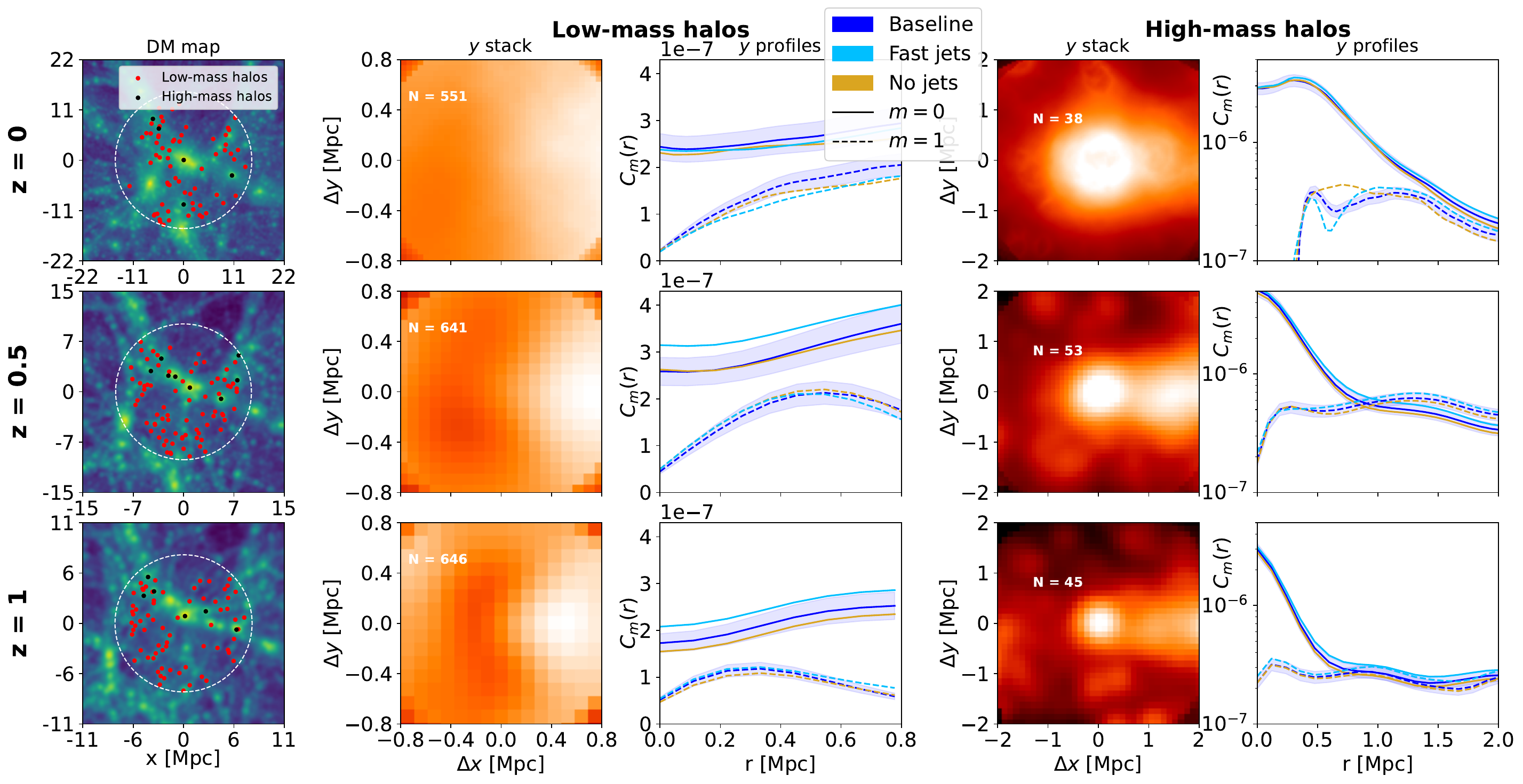}
    \caption{Stacked low-mass ($M<10^{13.5}$~\Msun, \textit{middle two panels}) and high-mass halos ($M>10^{13.5}$~\Msun, \textit{right two panels}) oriented by the dark matter field (shown at the far left) smoothed with a Gaussian of 0.5 and 5~Mpc FWHM, respectively. The $y$-maps are pixelated at SO pixel size but not smoothed with a beam. Radial profiles show the isotropic and dipole moments of the stacked images. Shaded bands show the estimated standard error on the mean for equivalent SO observations  for the baseline profile. The feedback effects are more impactful on the low-mass sample, and more strongly impact $m=0$ than $m=1$, but are all well within the estimated errorbars.}
    \label{fig:grid_all_z_stacks}
\end{figure*}

In this section, we use oriented stacking to investigate the shape and extent of the gas pressure distribution as a function of redshift. We determine the impact of feedback variations on these features by comparing the average oriented tSZ signal surrounding halos of a given mass range $[M_\mathrm{200,min}, M_\mathrm{200,max}]$. Observationally, the structure orientation would be typically estimated using the galaxy overdensity field; however, because we have access to the dark matter field we choose to use it to decrease the noise of the measurement. For each simulation region in the baseline feedback run, we first project the dark matter density along the $z$ axis, over 20 proper Mpc, to create a projected overdensity map $\delta(x,y)$. Next, we smooth this map with a Gaussian filter with full-width at half-maximum (FWHM) $R$ to create the smoothed map $F(R)$. We compute the first and second derivatives across the smoothed map, which will be used to identify features of the large-scale structure at selected points.

We select two main halo samples for stacking, a low-mass sample with $M\in[10^{12},10^{13.5}]$~\Msun\, and high-mass sample limited to $M>10^{13.5}$~\Msun. The low-mass sample encompasses massive galaxies and smaller galaxy groups, while the high-mass sample includes massive groups and clusters. We limit both samples to halos in the inner 15 $\mathrm{cMpc}$ of the snapshot to avoid edge effects. For the low-mass sample, we orient at small scales, setting $R=0.5$~Mpc, while $R=5$~Mpc for the high-mass sample. The dark matter $\delta$ map smoothed at the smaller scale, with the two samples overlaid, is shown for three redshifts in the left column of Figure~\ref{fig:grid_all_z_stacks}. 

For each sample, we take the projected $(x,y)$ positions of each halo and compose, at each position, the Hessian matrix from the second derivatives of the projected dark matter map:
\begin{equation} \label{eq:Hessian}
H = \begin{bmatrix}
\dfrac{\partial^2 F}{\partial x^2} & \dfrac{\partial^2 F}{\partial x\partial y} \\
\dfrac{\partial^2 F}{\partial y\partial x}  &\dfrac{\partial^2 F}{\partial y^2}
\end{bmatrix}
\end{equation}
The two eigenvectors of the Hessian identify the major and minor axes of the large-scale structure centered on each point.

Next, we construct small cutouts from the Compton-$y$ map centered on the halo positions. The map is the same as used in the previous section: noiseless, but with pixelization and smoothing consistent with expectations for SO, and also projected along the $z$ axis. We rotate each cutout by the angle between the long-axis eigenvector and the $x$ axis to align the structure horizontally. Additionally, we use the gradient of the overdensity map along the major axis to determine whether to flip the image after rotating -- if the gradient is positive in the $+x$ or $+y$ direction of the rotated image, i.e. the structure is becoming more dense along or perpendicular to the filament axis, the image is kept as-is. If it is becoming sparser, the image is flipped. Finally, the full sample of cutouts is stacked. This process places filamentary structure along the horizontal axis, and preferentially to the right and upwards, in the stacked image.

Due to the limited simulation sample size, we repeat this process for projections along the $x$ and $y$ axes of each snapshot, treating them as independent because they probe different projections of the same 3D structure. The stacks from the baseline $y$-map are shown in the second and fourth columns of Figure~\ref{fig:grid_all_z_stacks}. We then repeat the stacking procedure for the $y$-maps produced from the alternative feedback simulations, using the \textit{same} locations and orientations from the baseline run \rh{(defined by the matter density second derivatives)} in order to eliminate noise from variance in the halo sample when comparing stacks. Therefore, we examine the same exact structure in the different simulations, \textit{searching only for differences in the gas structure}. 
To compare the results quantitatively, we take the cosine decomposition of the stacked image $I$ as a function of distance from center $r$ and moment $m$:
\begin{equation}\label{eq:multipole_moments}
    C_m(r) = \frac{1}{X\pi}\int_0^{2\pi}  I(r,\theta)  \cos{(m \theta)} \dv{\theta},
\end{equation}
where $X=2$ for $m=0$ and $X=1$ otherwise, and $\theta$ is the polar angle measured counter-clockwise from the positive $x$-axis. There is weak signal in the analogous sine moments $S_m(r)$ as well, but it is small by construction of orienting structure along the $x-$axis, so we show only the cosine moments in the third and last columns of Figure~\ref{fig:grid_all_z_stacks}. 

The low-mass stack signal increases with $r$, because halos in this low-mass sample tend to be satellites of more massive halos with stronger tSZ signal. Meanwhile the high-mass stack shows a decline with $r$. There is a small decrement in the center of the high-mass stack at $z=0$, where the temperature is cooler \citep[for a previous study from \theth on cool-core clusters, see][]{Li2020MNRAS.495.2930L}. The dipole moment rises and falls according to the fixed physical smoothing scale and the alignment of structure.

We observe that the relationships between jet modes are redshift and multipole dependent. The jet variations cause subtle shape and amplitude impacts to the $m=0$ profiles (shown in solid curves in Figure~\ref{fig:grid_all_z_stacks}), which suggest that faster jets redistribute and heat gas further beyond the stacked clusters (right). The jets also increase the mean tSZ signal around low-mass halos (center) at $z=0.5$ and $z=1$, but have little impact on shape. At low-$z$, there is less impact to $m=0$. The $m=1$ signal is very subtly affected in shape, and we find that higher-order moments are even less impacted (not shown). We do not show results for the late coupling and random-direction runs in this figure due to being run only for a single region, but we find that the random-direction results most closely mimic the baseline jets, while the late coupling has diverse impacts depending on multipole and redshift, often falling between the fast jet and baseline results.

\begin{figure*}
        \centering
        \includegraphics[width=0.65\linewidth]{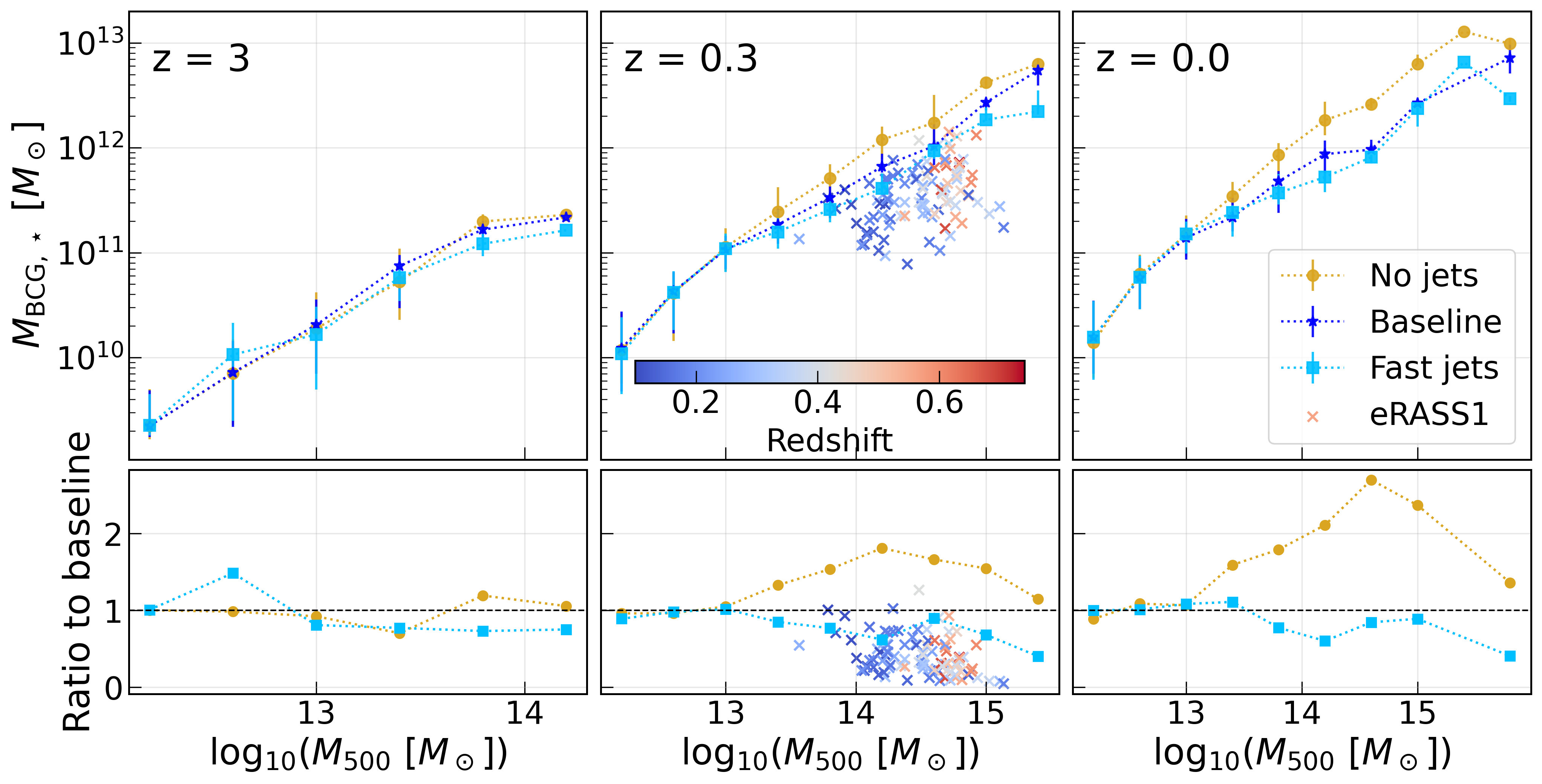}
        \caption{Stellar mass of the brightest cluster galaxy (BCG) as a function of halo mass ($M_{500}$) at $z = \{3, 0.3,0\}$ for three jet feedback models. In each halo mass bin, we compute the mean BCG stellar mass with 68\% confidence limits. On the $z = 0.3$ we overplot the eRASS1 catalog. The ``No jets'',``Baseline'', and ``Fast jets'' lines include data from all three simulation regions.}
\label{fig:bcg_mass}
\end{figure*} 

It is challenging to estimate realistic uncertainties on the profiles, as the cutouts in the stacked samples are not independent --- along a given projection direction they often overlap, sampling similar structure, and the cutouts along different projections of the same snapshot sample the same 3D structure. However, as a reference, we can look to the literature to estimate errors on a realistic analysis and ask whether it is likely to be able to distinguish the feedback differences. In stacking ACT $y$-maps with orientation on $M>10^{14}$~\Msun\; halos, \citet{Lokken_2025} measured extended signal ($>5$~Mpc) at an uncertainty level of $\sim25\%$ for bins of width $\Delta z \sim 0.15$. This level of uncertainty would make the profiles shown in Fig.~\ref{fig:grid_all_z_stacks} indistinguishable. In the lower-mass regime, where the $y$ signal is smaller but halos are more abundantly available for stacking, \citet{Tanimura2019MNRAS.483..223T} measured filament gas to $\sim20\%$ precision using \textit{Planck} data. However, the SO $y$-maps are expected to have $\sim2-3$ times smaller noise power at the relevant scales ($\ell\sim1000-2000$) than ACT \citep{coulton2023atacamacosmologytelescopehighresolution, thesimonsobservatorycollaboration2025simonsobservatorysciencegoals}. With $\sqrt{2}$ to $\sqrt{3}$ times smaller errorbars in the stacked signal (which contains only one factor of the map, as opposed to the noise power spectrum), the largest differences between feedback models in both the low-mass and high-mass end would become marginally distinguishable. We show uncertainties of $\pm0.2\,C_m(r)/\sqrt{3}$, where $C_m(r)$ is the value of each moment, in faint shaded bands around the baseline curves in Fig.~\ref{fig:grid_all_z_stacks} to represent this rough reference. Such analyses can also gain statistical power in the future by using, for the stacking points, overlapping galaxy surveys like Euclid \citep{Euclid2022A&A...662A.112E} and Rubin LSST \citep{LSST2019ApJ...873..111I}, which will have deeper observations and smaller redshift uncertainties than the previous generation of surveys used for past analyses. However, the anlayses are limited by contamination by the cosmic infrared background and radio sources, which remain a poorly-understood systematic \citep[see, e.g.,][for more discussion]{Liu2025PhRvD.112h3561L} that can inflate errors by factors of two or more.

Making robust feedback constraints from oriented stacks using observational data will also be challenging from the modeling standpoint. For direct comparison of simulations to data, much larger volumes are necessary; the Flamingo simulations \citep{Schaye2023MNRAS.526.4978S} that have recently been released, along with sets of mock CMB secondary maps, present a new opportunity. However, accurate comparisons require replicating the halo population of the observations in simulations, which is challenging. Therefore, it is unclear whether tSZ analyses with SO will be able to rule out particular phenomology of jet feedback.

Additionally, the absolute amplitude of the $m=0$ profiles of a stack is difficult to measure. Because instruments like ACT and SO do not measure a meaningful $y$-mean, and because there is large-scale contamination from CMB modes in the maps which cause large fluctuations in the background of stacks, only the differences in profile shape can be robustly constrained. Thus, the amplitude differences in, for example, the ($z=0.5$, $m=0$) profile are unlikely to be measurable, as the scales are also too small to be measurable by lower-resolution proposed spectral distortions instruments like FOSSIL.

Lastly, we note again that there is no reason to expect these results (stacks on halos in and around massive clusters and actively-forming protoclusters) to be consistent with effects of feedback in less dense regions of the cosmic web. These results show that the massive-halo regime, especially in higher-order moments of stacks, are not highly sensitive to jet feedback differences. Stacks on similar-mass halo samples, or filaments, in more isolated regions of the cosmic web might be more sensitive to feedback differences; this has been recently explored in \citep{Hadzhiyska2025PhRvD.112h3509H, hadzhiyska2025PhRvD.111b3534H}.

\subsection{Brightest cluster galaxies}
Brightest cluster galaxies (BCGs) are luminous galaxies located toward the center of their cluster. They are typically old, elliptical and are some of the brightest galaxies in the universe. They are of particular interest because it is believed that their formation is linked to that of their host cluster \citep{Lin_2004}. It has now been widely observed that the cooling of gas towards the center of BCGs is suppressed by AGN feedback \citep{McNamara_2007, Gitti_2012, Fabian_2012}. For this reason, the stellar mass of BCGs is used as an observational diagnostic for testing and calibrating AGN feedback models in cosmological simulations \citep{Cui_2022}. We examine how the stellar mass of the BCG scales with its host halo mass across the feedback variations. We examine this for three redshifts: $z = \{3,0.3,0\}$, allowing us to assess both the early assembly of BCGs and their late-time evolution. 

We present the results in Figure~\ref{fig:bcg_mass}, which shows the stellar mass of BCG galaxies as a function of halo mass for the varying feedback runs. In the $z = 0.3$ panel we overlay the BCG mass-halo mass relation with the eROSITA All-Sky Survey cluster catalogs \citep[eRASS1;][]{bulbul2024srgerositaallskysurveycatalog}, which provide the largest sample of X-ray–selected galaxy groups and clusters to date. The BCGs from eRASS1 spans $z \leq 0.74$ with a median redshift of $z=0.29$. BCGs were identified for a subset of this data during optical follow-up of the X-ray catalog; each BCG is the brightest member of its host cluster in the $z$-band \citep{erass_optical_2024A&A...688A.210K}. For 101 of these BCGs, both their stellar masses and the corresponding cluster halo masses have been estimated using Hyper Suprime-Cam data \citep[SED-fitting for the stellar mass, and weak-lensing calibration for the halo mass, see][]{bcg_masses_erass2025A&A...704A.110C}.

Figure~\ref{fig:bcg_mass} shows that higher jet velocities are more effective at expelling gas from the central regions and heating its surroundings \citep{Martizzi2012}. This suppresses cooling flows, resulting in lower BCG stellar masses. The comparison with eRASS1 indicates that the higher-velocity jet models provide a better match to the observed normalization of the BCG stellar mass-halo mass relation, particularly at intermediate halo masses\footnote{Although we repeat the caution against drawing too-strong conclusions based on simulations of three cluster regions and the resolution limits.}. While we only look at varying AGN jets in our simulations, BCG stellar mass may also be very sensitive to wind-mode feedback. AGN winds are typically isotropic and have lower velocities. Consequently, they are efficient at heating the regions near where they are emitted and hence preventing sufficient cooling for star formation. 

An issue with BCGs in hydrodynamic simulations is that it is difficult to separate them from the intracluster light \citep[ICL, see][]{Dolag_2010,Canas_2020}. As a result, comparing them directly with observations is not a simple task \citep{Cui_2013}. Previous analyses within \theth have proposed calculating the BCG stellar mass by summing all the stellar mass within a fixed aperture of the halo center. \cite{Cui_2022} used apertures of $30\, \mathrm{kpc}$, $50\, \mathrm{kpc}$ and $0.1 R_{500}\,\mathrm{kpc} $ respectively, while \cite{Contreras_Santos_2022} used $30\,\mathrm{kpc}, 50\,\mathrm{kpc}$ and $70\, \mathrm{kpc}$. We define the BCG using an aperture of $30\, \mathrm{kpc}$. However, changing the radius of the aperture only affected the stellar mass on the $\sim20\%$ level, which did not affect its agreement with eRASS1. Additionally, we calculated the stellar mass using the \caesar $\texttt{central\_galaxy}$ tag to identify the BCG, which also yielded similar results at low redshift. See \cite{Cui_2022} for a comparison of BCG aperture and stellar mass within \theth. Regardless of how we define the BCG, our results suggest that relatively strong AGN feedback is required to prevent the overproduction of stars in massive central galaxies. 
\vspace{-0.1in}
\subsection{Black hole mass}
\begin{figure}
    \centering
    \includegraphics[width=0.75\linewidth]{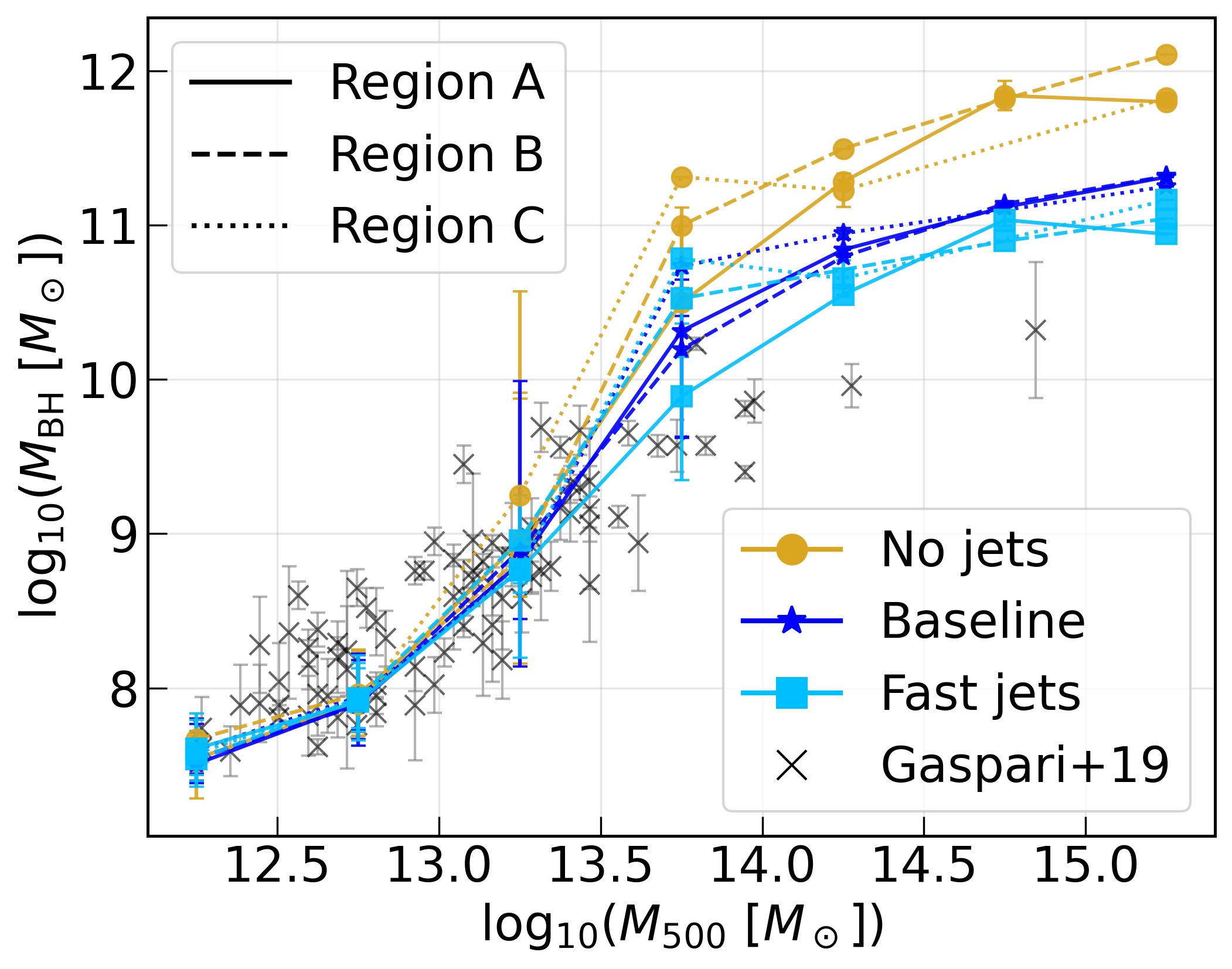}
    \caption{Total BH mass in the host halo as a function of halo mass at redshift $z = 0$. Each of the simulation runs are plotted as different lines where the errorbar is the 68\% confidence limits. Each bin has a horizontal width of 0.5 $\mathrm{log}_{10}(M_{500} [M_\odot])$. The different colors correspond to different jet velocity runs, whereas the different line styles correspond to different cluster regions. The black crosses overlaid are observations from \cite{Gaspari_2019}. }
    \label{fig:bh_halo_mass}
\end{figure} 
\vspace{-0.1in}

\cite{Gaspari_2019} studied the properties of 85 galaxies, groups and clusters that had been observed in both the optical and X-ray bands. The BH masses are found in \cite{Bosch_2016} while the the halo properties are taken from Chandra, ROSAT, and XMM-Newton \citep{Sullivan_2003, Diehl_2008,Nagino_2009, Kim_2015, Su_2015,Goulding_2016}. \cite{Gaspari_2019} provide more details on the selection methodology and a calculation of halo properties. The measurements we use in this work are their measurements of the BH mass $M_{\bullet}$ and $R_{500},$ defined as the radius at which the mass enclosed, $M_{500},$ has a mean mass density 500 times the critical density $\rho_c$ of the universe (see Equation \ref{eq:critical_density}). We acquire $M_{500}$ then as $
M_{500} = \frac{4\pi \rho_c 500 R_{500}^3}{3}.$ \begin{figure*}[t]
    \centering
    \includegraphics[width=0.92\linewidth]{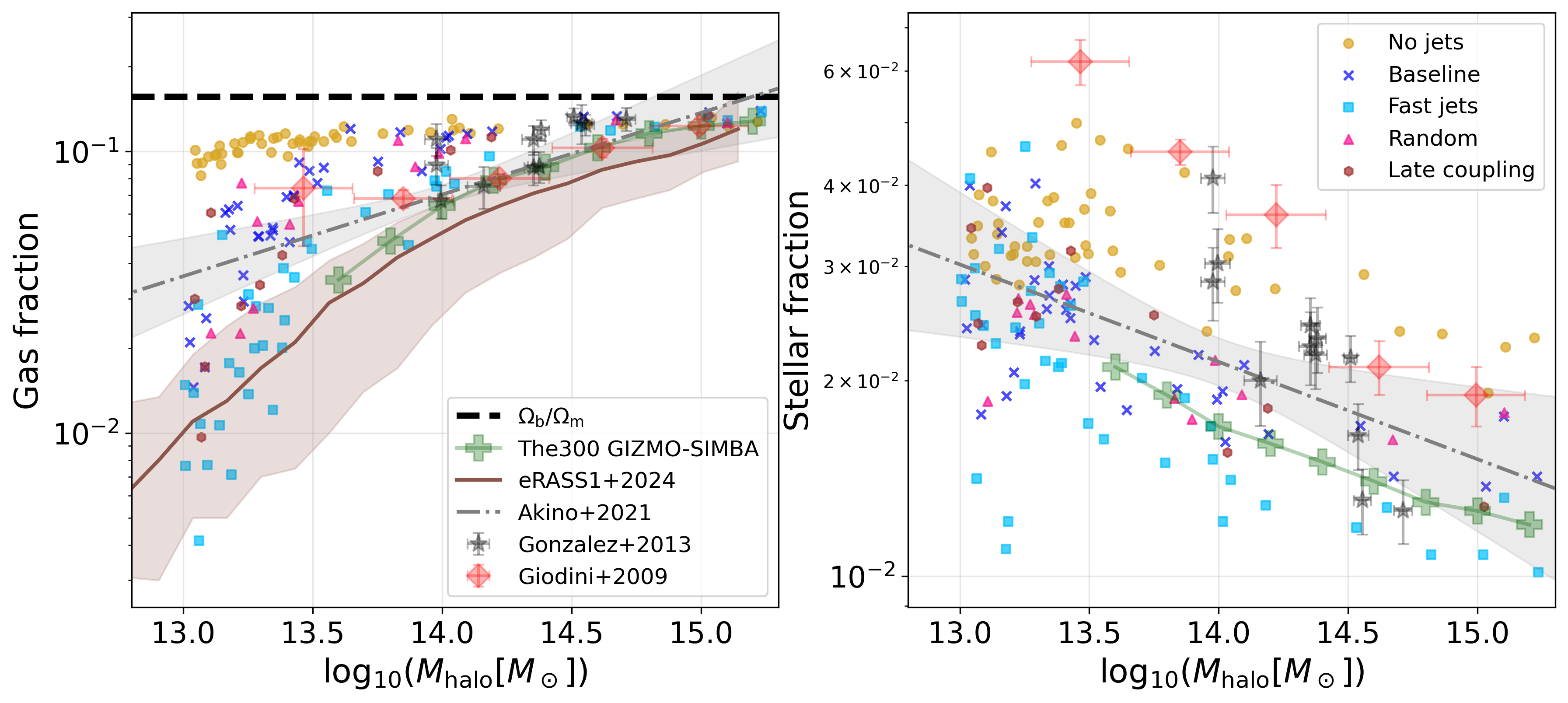}
    \caption{Left: Gas mass fraction within $R_{500}$ in the host halo (for halos within $15\, h^{-1}\mathrm{cMpc}$) as a function of the halo mass ($M_{500})$ at redshift $z = 0$. All of the simulation regions for a given jet velocity are combined. The different colours correspond to different jet feedback runs, and the dotted black line is the cosmic baryon fraction. Right: Same as the left but for stellar mass fraction. Four sets of observations are overlaid for the gas fractions: \cite{Giodini_2009}, \cite{Gonzalez_2013}, \cite{Akino_2022} and \cite{bulbul2024srgerositaallskysurveycatalog}. All but the latter are also overlaid with the stellar mass fractions. We additonally overlay the gas and stellar fractions the \gizmo-\simba run of \theth using all 324 regions \citep{Cui_2022}.}
    \label{fig:baryonic_mass_fraction}
\end{figure*} 
We examine the relation between total BH mass and halo mass, shown in Figure~\ref{fig:bh_halo_mass}, using halos within $15 \, h^{-1}\, \mathrm{cMpc}$ from the center of the simulation volume, for all three simulation regions. This relation is approximately linear in log–log space, which is consistent with previous simulations \citep{Cui_2022}. Previous work has shown that BH mass is correlated with gas mass, which is strongly affected by AGN feedback \citep{Gaspari_2019}. We find that for lower mass halos, the three runs are all in alignment with each other, and with observations. This suggests that in this regime BH growth is regulated primarily by the \textit{global halo gas supply rather than the detailed implementation of jet feedback}. The linear relation between BH and halo mass appears to flatten around $M_{500} = 10^{14} M_{\odot}$. In these systems, efficient AGN feedback suppresses gas cooling and star formation at early times, leading to continued halo growth through mergers and accretion without a corresponding increase in BH mass \citep{Cui_2022}. In general, increasing the jet velocity leads to a reduction in the total BH mass at fixed halo mass. This shows that more powerful jets more efficiently heat and expel gas from the central regions, suppressing accretion onto the BH and limiting its subsequent growth. Overall, the agreement between the simulations and observational constraints from the scale of massive galaxies to low-mass clusters, combined with the systematic trends observed at the high-mass end, supports the interpretation that AGN jet feedback plays a central role in regulating BH growth by controlling the thermodynamic state of the halo gas. We leave the analysis of the resolution effects of the BH mass estimation to future work. As mentioned in the section above, BH growth is also very sensitive to BH winds, which would also heat the gas around the BH to prevent cooling inflows.
\vspace{-0.1in}

\subsection{Baryonic mass}
AGN feedback is efficient in expelling gas from halos and quenching star formation \citep{CanoDiaz2012A&A...537L...8C, Schaye_2014, sorini2022MNRAS.516..883S}.
The \simba model has proven to be particularly effective at evacuating baryons from group-sized halos compared to many other simulations \citep{Oppenheimer_2021}, which is more consistent with recent measurements from eROSITA \citep{Popesso_2026}.
In Figure~\ref{fig:baryonic_mass_fraction}, we plot the gas and stellar mass fractions for halos at redshift $z=0$. We find that AGN feedback strongly suppresses both components across the halo mass range, with the largest relative impact occurring in lower-mass systems.

We see the expected trend that the stellar mass fraction declines with halo mass, while the gas mass fraction increases toward higher-mass systems. \citet{Giodini_2009} studied galaxy groups and clusters from the COSMOS 2~deg$^2$ survey over the mass range $10^{13}M_\odot \lesssim M_{500} \lesssim10^{15}M_\odot$, and found that the stellar fraction decreases steeply with increasing halo mass, while the total baryon fraction increases but remains systematically below the cosmic value. Even after accounting for plausible contributions from ICL and gas depletion during hierarchical assembly, they find a significant baryon deficit at the group scale, which they interpret as evidence for non-gravitational processes such as feedback and filamentary heating. \cite{Gonzalez_2013} similarly find that the total mass of stars and X-ray gas are strong functions of halo mass. A key result of that work is the importance of the combined BCG and intracluster light contribution to the stellar mass budget, particularly in lower-mass halos. Studies that neglect the intracluster stellar component tend to underestimate the stellar fraction by $\sim 25\%$. \cite{Akino_2022} study an X-ray–selected XXL sample of 136 galaxy groups and clusters spanning $M_{500} \sim 10^{13}$–$10^{15},M_\odot$ and $0 \lesssim z \lesssim 1$. Using weak-lensing–calibrated halo masses, they find a steep gas mass–halo mass relation and a comparatively shallow stellar mass–halo mass relation. 

The stellar mass fractions inferred by \cite{Akino_2022} differ in normalization and slope from \cite{Gonzalez_2013} and \cite{Giodini_2009}. However, this difference reflects the broader diversity of methodologies in the literature rather than a clear inconsistency, as these studies differ in cluster selection (X-ray versus optical), halo mass calibration, and in how mass bias and selection effects are modeled. 

We also compare our results to constraints from the eRASS1 cluster catalog \citep{bulbul2024srgerositaallskysurveycatalog}. The halo mass is not measured directly, but inferred through a cosmology pipeline that links the observed X-ray count rate to halo mass using a scaling relation calibrated with weak-lensing shear measurements \citep{Ghirardini_2024}. The catalog contains 12,247 optically confirmed galaxy groups and clusters detected in the 0.2-2.3 ~keV range, of which 10,440 have estimates of gas fractions. 

In Figure \ref{fig:baryonic_mass_fraction} we overlay the observational campaigns mentioned above with data for all the jet parameter variations. For the gas fraction from the eRASS1 data, we divide the catalog into 20 bins equal in log halo mass and compute the median and 84\% confidence interval, shown in the brown shaded region on the left panel. From our simulations, we only include halos with masses above $10^{13}\, \mathrm{M_{\odot}}$ to ensure the gas and stellar mass fractions are not subject to resolution limits. We find that AGN feedback is particularly efficient at removing gas from lower mass halos due to their shallower potential wells, with the displaced gas preferentially settling at larger radii. This dip in gas fraction at group scale masses also appears in the eRASS1 catalog. This provides very strong evidence for the existence of strong AGN feedback in halos with $M_{500} \sim10^{13-13.5} \mathrm{M_{\odot}}$.

For stellar mass fraction, we do not see as much of a change between the feedback modes for group sized halos. This is because by the time the jets turn on (after the black hole has reached $M_\bullet = 10^{7.5}$), most of the stars have already formed. From Figure \ref{fig:jet_props} we see that the maximum number of active jets occurs around $z\sim 1-2$, at which point the star formation rate has already peaked. So, AGN jets are efficient at heating and expelling gas but are too late to completely suppress stellar mass. Still, AGN feedback is efficient in lowering the stellar mass fraction of halos, from the group to cluster scales. See \cite{Ma2025} for a greater discussion on the effects of AGN feedback on stellar mass within the \simba simulation. 

In contrast, in simulations \textit{without AGN jets}, the total gas mass within halos remains approximately \textit{constant as a function of halo mass}. We find that the jet emission direction and decoupling time do not have a strong effect on the gas and stellar fractions. Our results are largely consistent with previous work, which finds that halos retain $90 \pm 6\%$ of the cosmic baryon fraction within their virial radius \citep{Crain_2007}. 

Lastly, we compare our runs with the gas and stellar fractions from the \gizmo-\simba run of \theth which are calculated across all 324 regions at the same resolution as our work \citep{Cui_2022}. We omit its errorbars for clarity but for halos of mass $10^{13.5}\,\mathrm{M_{\odot}}$ they approach order unity. Their modeling uses a maximum jet velocity of $15,000~\mathrm{km~s^{-1}}$ and is in better alignment with our stronger feedback runs. The primary differences between our simulations and the \gizmo-\simba run are likely driven by our smaller selection of regions and our use of \simbac rather than \simba. 

\subsection{Random direction jets}

We examine the impact of randomly-orienting the jets, instead of aligning them with the black hole angular momentum. By fixing the jet direction there is a gradual build-up in AGN feedback across cosmic time. This may allow for hot gas from jets to permeate further out of its host cluster. Under a scenario with randomized jet directions, the heating is expected to be more isotropic. 

When comparing the temperature and Compton-$y$ fields, we do not observe such a coherent effect. In general, the random orientation leads to cooler gas on large scales. However, there could be cases where randomly orienting the jets could let them travel further, if they are ejected in a direction with lower pressure media. Overall, changing the jet direction appears to have a complicated, non-uniform effect on the thermal state of the gas; we leave it for further studies.

In motivating this run, we also considered that the direction of a galaxy's AGN jet emission may be aligned with its galactic neighbours. Previous simulation work has shown that the spin of lower-mass blue galaxies are preferentially aligned with their neighbouring filaments, while higher-mass red galaxies tend to have peculiar spins \citep{Dubois_2014} (this may be driven by past mergers which cause spin misalignment). Therefore, it is possible that alignment versus randomization of the jet direction would affect the morphology of the filaments.

We do not observe alignment with large-scale structure in the differences between the random-direction and baseline run. In fact, near the filament axes, the differences appear to be minimal, while they are most impactful in underdense regions. This finding was corroborated by examining this run with the other statistical methods used in this work, finding in all cases that the impacts were less significant than for the case of increased jet velocity. The lack of effect may be explained by the fact that, from Equation~\ref{eq:jet_vel}, we see that the most powerful jets are associated with the most massive black holes. As a result, even if smaller galaxies tend to be aligned with one another, the feedback that we see in our simulations may be dominated by a few very large galaxies that are less aligned with the large-scale structure.

We perform the same analysis as described in Section \ref{ssec:overdensity} for our random direction and late coupling runs, shown in Figure \ref{fig:decoupling_overdensity}. In the top row, we see the gas mass above $10^7 \, \mathrm{K}$ for the random direction and late coupling runs, respectively, divided by that of the baseline run. The bottom row is the same but for gas between $10^5-10^7 \mathrm{K}$. We find that randomizing the jet emission direction reduces the gas mass above $T>10^7\,\mathrm{K}$ in underdense regions by nearly $50\%$, which is persistent past redshift $z = 1$. This supports the idea that the continual build up of feedback enables hot gas to travel long distances. As in Figure \ref{fig:gas_frac_z}, we only plot the mass when at least 100 particles are found in that density range at that redshift.

\subsection{Late coupling time jets}\label{ssec:decoup}

Under the late coupling model, jets can travel up to $\sim 200 \,\mathrm{kpc}$ at $z=0$ before recoupling, as opposed to $\sim 10\, \mathrm{kpc}$. Not only does this change the distance that the jets travel, but also prevents them from heating and pushing gas in the inner regions from where they are ejected.

\begin{figure}[]
\centering
\includegraphics[width=0.99\linewidth]{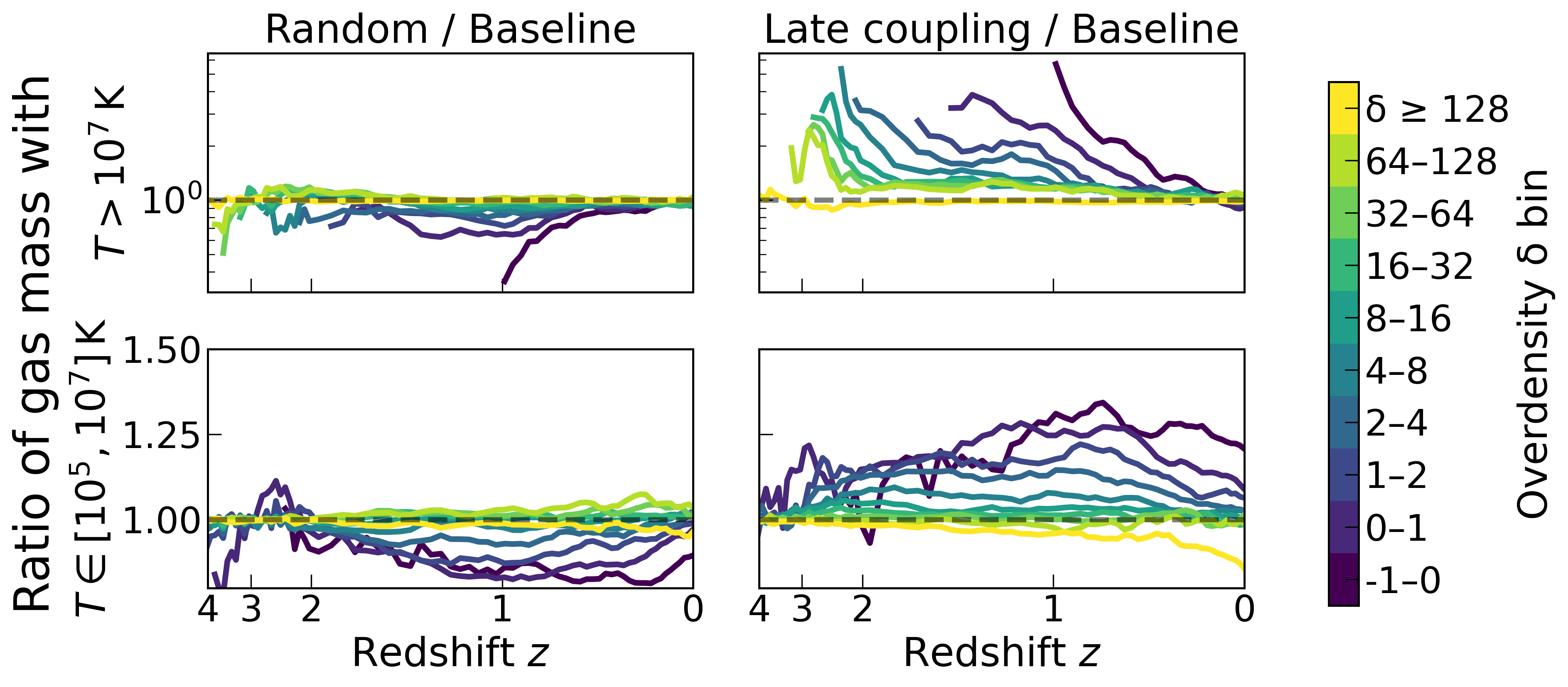}
        \caption{The gas mass above $T>10^7\, \mathrm{K}$ (top row) and between $10^5 - 10^7\, \mathrm{K}$ (bottom row) for different feedback runs divided by the baseline run, with separate lines for different overdensity values.
        An example of the separation of the gas into density groups is shown in Figure~\ref{fig:gas_frac_z}. The left column shows the random direction run divided by the baseline, while the right column is the late coupling divided by the baseline. Each of the colors corresponds to gas in different overdensity regimes.}
    \label{fig:decoupling_overdensity}
\end{figure}

Figure \ref{fig:decoupling_overdensity} shows that this too has an effect on the gas temperature in the most underdense regions. This is most impactful around redshifts $1 - 4$ where the later coupling time yields $\sim 5$ times more gas above $10^7\, \mathrm{K}$ in the least dense regions. At higher redshifts, there are fewer jets being emitted, and the coupling time is shorter (due to proportionality with the Hubble time) so we may naively expect not to see as much of a difference in temperature. However, since clusters may only be a few hundred $\mathrm{kpc}$ across, if there are any decoupled jets travelling at their maximum velocity, they would be able to exit the cluster much easier. By redshift $z \sim 0$, the jets with later coupling are traveling up to the entire $R_{500}$ radius before recoupling. The changes the decoupling time makes to gas in the $10^5\,\mathrm{K}  <T<10^7\,\mathrm{K}$ regime is relatively smaller. We note that it appears in the densest regions that the hot gas mass for the late coupling dips below the baseline at $z = 0$. It is unclear if this is physical or random noise; we leave this for future work.

\section{Conclusions}\label{sec:conclusion}
We investigated how varying the implementation of AGN jet feedback affects the thermodynamic state of baryons and their observable signatures on $\sim$Mpc scales. Using zoom-in hydrodynamic simulations of three \theth proto-cluster regions with \simbac, we varied the jet velocity cap, the jet emission direction, and the jet decoupling time. We quantified changes in the IGM/WHIM temperature distribution, halo baryon content, BH and BCG growth, and the tSZ signal in and around clusters. Our goal was to move beyond ``jets on/off'' comparisons and instead map out how subtler changes in jet parameters translate into differences in the WHIM, galaxy properties, and tSZ observables. Our main conclusions are as follows:
\begin{itemize}
    \item \textbf{AGN jets primarily heat diffuse environments rather than dense clusters.} When we separate the gas by overdensity and track the mass fraction above $T>10^7$~K (Figure~\ref{fig:gas_frac_z}), we find that differences between feedback models are modest in the highest-density regimes associated with clusters and filaments, however, the \textit{least dense environments show strong sensitivity to jet prescriptions}. In particular, increasing jet velocity produces nearly an order of magnitude more $T>10^7$~K gas at intermediate redshifts ($z\sim1-4$) in low-density regions, indicating that powerful jets are efficient at transporting energy into the IGM.

    \item \textbf{Hot gas in underdense regions is a distinctive signature of jet feedback.}
    Across our runs, the presence (or absence) of very hot gas in the most underdense regions provides a clear discriminator between jet models. The fast jet model populates diffuse environments with $T>10^7$~K gas \textit{earlier and more strongly} than the baseline, while the no jet run produces markedly less hot diffuse gas (see Figures~\ref{fig:igm_temp} and \ref{fig:gas_frac_z}). We note that since our zoom regions are centered on massive clusters, these ``void'' environments are not necessarily comparable with cosmic voids. Nevertheless, \textit{the appearance of hot gas in underdense regions appears to be a strong indicator of efficient AGN-driven energy transport}.

     \item \textbf{tSZ profiles show the largest fractional feedback differences outside filaments, but these are observationally challenging.}
     By separating pixels into filamentary and non-filamentary regions using \textsc{DisPerSE} skeletons (Section~\ref{ssec:dispere}), we find that stronger feedback enhances Compton-$y$ both inside and outside filaments (Figure~\ref{fig:sz_fil}). The relative differences are typically $\sim10\%$ inside filaments but can approach order unity outside filaments, where the absolute signal is faint. This suggests that \textit{stacked analyses will be essential} for extracting these signatures from future CMB surveys.

     \item \textbf{The jet impacts on stacked tSZ profiles are weak and depend on mass and redshift.}
     Using oriented stacking tied to the smoothed dark matter field, we find that variations in jet strength  are redshift and multipole dependent. The jet variations affect the mean-$y$, shifting the amplitude of $m=0$ profiles, and in most cases the fast jets mildly flatten the pressure profile (Figure~\ref{fig:grid_all_z_stacks}). The weak impacts compared to predicted errorbars suggest that oriented signal in the near-cluster regime is relatively robust to feedback uncertainties.
    \item \textbf{AGN jets suppress central galaxy growth and black hole growth in massive halos, with observations consistent with stronger feedback.}
    Increasing jet strength reduces BCG stellar masses at fixed halo mass (Figure~\ref{fig:bcg_mass}) and lowers the total BH mass in the most massive systems (Figure~\ref{fig:bh_halo_mass}). These trends support the physical picture in which more \textit{powerful jets heat and evacuate central gas}, suppressing cooling flows and limiting both star formation in BCGs and black hole accretion. Through comparison with X-ray observations, the faster jet run provides the best match of black hole and BCG mass for halos on the $10^{14} - 10^{15} M_{\odot}$ scale. The magnitude of these effects may be sensitive to resolution.
    
    \item \textbf{AGN jet feedback strongly suppresses halo baryon and stellar mass fractions, with the largest impact at group scales.}
    We find that AGN feedback significantly reduces both the gas and stellar mass fractions within $R_{500}$ at redshift $z=0$ (Figure~\ref{fig:baryonic_mass_fraction}). The effect is strongest in lower-mass halos ($10^{13}M_\odot \lesssim M_{500} \lesssim10^{14}M_\odot$), where shallow potential wells make baryons susceptible to removal and redistribution by feedback. In these systems, the fast jet run leads to a baryon fraction \textit{nearly an order of magnitude lower than the baseline and no jet runs}. The fast jet run provides the best match to the gas fractions observed in the eRASS1 catalog.

    \item \textbf{Jet orientation affects the gas temperature in a non-uniform way.}
    Qualitatively, randomized jets often yield cooler gas relative to the angular-momentum-aligned case, consistent with the idea that \textit{repeated energy injection along preferred directions are needed to allow propagation to Mpc scales} (see Figure~\ref{fig:decoupling_overdensity}). The randomized-direction model does not strongly modify $z=0$ black hole masses, baryonic fractions or BCG stellar masses relative to aligned jets.

    \item \textbf{Varying the jet decoupling time allows heating of regions further from clusters.}
    Extending the decoupling time allows jets to travel substantially farther before interacting with the surrounding gas. The jets with \textit{later coupling time can reach underdense regions at earlier times}, leading to more hot gas at $z\sim 1-4$, see Figure~\ref{fig:decoupling_overdensity}. We do not find as strong shifts compared to changes driven by velocity $v_{\rm Max}$. 
    
\end{itemize}

These results emphasize that the most sensitive large-scale signatures of AGN jet modeling appear in the low-density IGM and in cluster outskirts, rather than in the densest nodes of the cosmic web. In particular, the emergence of very hot gas in underdense regions at $z\gtrsim1$ is a promising observational discriminant that can be targeted with next-generation CMB experiments via stacked tSZ measurements. This work represents a first step in systematically mapping jet-parameter space within a fixed hydrodynamic framework. Future extensions include: (i) repeating these parameter variations across a much larger subset (ideally all) of the \theth regions; (ii) running larger box simulations to capture a more representative population of halos; (iii) applying filament-finding directly in 3D galaxy fields to isolate morphological changes; (iv) performing oriented stacking of clusters along the axis of jet emission to understand how they are reshaping gas along their path; and (v) producing mock-observations (including noise) to determine which of the signatures identified here are realistically detectable with upcoming surveys such as SO and FOSSIL.

\begin{acknowledgements}
This work has been made possible by the ‘The Three Hundred’ collaboration, which has received financial support from the European
Union’s Horizon 2020 Research and Innovation programme under the Marie Sklodowska-Curie grant agreement number 734374, i.e. the LACEGAL project. Simulations in this paper were performed on the MareNostrum Supercomputer at the Barcelona Supercomputing centre, thanks to CPU time granted by the Red Espa\~nola de Supercomputaci\'on. The CosmoSim database used in this paper is a service by the Leibniz-Institute for Astrophysics Potsdam (AIP). The MultiDark database was developed in cooperation with the Spanish MultiDark Consolider Project CSD2009-00064. FAE is partially funded by the CERCA program of the Generalitat de Catalunya. This research was partially funded by the Spanish Ministry of Science and Innovation/State Research Agency (AEI) through a Juan de la Cierva postdoctoral grant [JDC2024-XXXXXX-I]. IR acknowledges financial support from the European Union’s Erasmus+ programme through the Erasmus Mundus Joint Master in Astrophysics and Space Science. We acknowledge the computing resources of the CITA cluster. Funding was provided by grants from the Natural Sciences and Engineering Research Council of Canada (RGPIN-2025-06483 and SMFSU-60768) the Connaught Fund. The Dunlap Institute is funded through an endowment established by the David Dunlap family and the University of Toronto. The authors at the University of Toronto acknowledge that the land on which the University of Toronto is built is the traditional territory of the Wendat Nation, the Seneca, and the Mississaugas of the Credit. Today, this meeting place is still the home to many Indigenous people from across Turtle Island. They are grateful to have the opportunity to work on this land.
\end{acknowledgements}

\bibliography{bib}{}
\bibliographystyle{aa}

\end{document}